\documentclass[%
 aip,
 amsmath,amssymb,
 reprint,%
]{revtex4-1}

\usepackage{graphicx}
\usepackage{dcolumn}
\usepackage{bm}

\usepackage[utf8]{inputenc}
\usepackage[T1]{fontenc}
\usepackage{mathptmx}
\usepackage{float}

\usepackage[dvipsnames]{xcolor}

\newcommand*{\dd}{\mathrm d}
\newcommand*{\ww}{\mathrm w}
\newcommand*{\hh}{\mathrm h}
\newcommand*{\aaa}{\mathrm a}
\newcommand*{\sGL}{\sigma_{\scriptscriptstyle\rm GL}}
\newcommand*{\sSL}{\sigma_{\scriptscriptstyle\rm SL}}
\newcommand*{\sSG}{\sigma_{\scriptscriptstyle\rm SG}}

\newcommand*{\sGW}{\sigma_{\scriptscriptstyle\rm GW}}
\newcommand*{\sGO}{\sigma_{\scriptscriptstyle\rm GO}}
\newcommand*{\sSW}{\sigma_{\scriptscriptstyle\rm SW}}
\newcommand*{\sSO}{\sigma_{\scriptscriptstyle\rm SO}}
\newcommand*{\sWO}{\sigma_{\scriptscriptstyle\rm WO}}
\newcommand*{\NvizW}{{\rm n}_{\scriptscriptstyle \rm W}^{\scriptscriptstyle \rm {i}}}
\newcommand*{\NvizG}{{\rm n}_{\scriptscriptstyle \rm G}^{\scriptscriptstyle \rm {i}}}
\newcommand*{\NvizO}{{\rm n}_{\scriptscriptstyle \rm O}^{\scriptscriptstyle \rm {i}}}

\newcommand*{\Scap}{S_{\scriptscriptstyle\rm CAP}}
\newcommand*{\ScapCB}{S_{\scriptscriptstyle\rm CAP}^{\rm CB}}
\newcommand*{\ScapW}{S_{\scriptscriptstyle\rm CAP}^{\rm W}}
\newcommand*{\ttc}{\theta_{\scriptscriptstyle\rm C}}
\newcommand*{\NW}{{\rm N}^{\rm W}}
\newcommand*{\NCB}{{\rm N}^{\rm CB}}
\newcommand*{\Ns}{{\rm N}^{\rm s}}
\usepackage[normalem]{ulem}

\begin{document}

\preprint{AIP/123-QED}

\title{Modeling oil-water separation with controlled wetting properties}

\author{Cristina Gavazzoni} 
\email{crisgava@gmail.com}
\affiliation{Instituto de F\'isica, Universidade Federal
do Rio Grande do Sul, Caixa Postal 15051, CEP 91501-970, 
Porto Alegre, Rio Grande do Sul, Brazil}

\author{Marion Silvestrini} 
\affiliation{Instituto de F\'isica, Universidade Federal
do Rio Grande do Sul, Caixa Postal 15051, CEP 91501-970, 
Porto Alegre, Rio Grande do Sul, Brazil}

\author{Carolina Brito}
\affiliation{Instituto de F\'isica, Universidade Federal
do Rio Grande do Sul, Caixa Postal 15051, CEP 91501-970, 
Porto Alegre, Rio Grande do Sul, Brazil}

\date{\today}

\begin{abstract}

Several oil-water separation techniques have been proposed to improve the capacity of cleaning water. With the technological possibility of producing materials with antagonist wetting behavior, as for example a substrate that repeal water and absorb oil, the understanding of the properties that control this selective capacity has increased with the goal of being used as mechanism to separate mixed liquids. Besides the experimental advance in this field, less is known from the theoretical side. In this work we propose a theoretical model to predict the wetting properties of a given substrate and introduce simulations with a 4-spins cellular Potts model to study its efficiency in separating water from oil. Our results show that the efficiency of the substrates depends both on the interaction between the liquids and on the wetting behavior of the substrates itself. The water behavior of the droplet composed of both liquids is roughly controlled by the hidrophobicity of the substrate.  To predict the oil behavior, however, is more complex because the substrate being oleophilic  do not guarantee that the total amount of oil present on the droplet will be absorbed by the substrate. For both types of substrates considered is this work, pillared and porous with a reservoir, there is always an amount of reminiscent oil on the droplet which is not absorbed by the substrate due to the interaction with the water and the gas. Both theoretical and numerical models can be easily modified to analyse other types of substrates and liquids.

\end{abstract}

\maketitle

\section{Introduction}

Water separation and purification methods have been widely studied due to environmental, economical and social issues \cite{mek16,huang2019fabrication,yang2019large,qian2019tuning}. Particularly, there is an increase level of attention focused in oil-water separation techniques mainly due to oil  been the most common pollutant in the world, principally from oil spill accidents and industry oily wastewater~\cite{xue14,chan09, pa15}.

Oil/water mixtures can be classified in two different ways depending on the diameter $d$ of the dispersed phase: stratified oil/water ($d>20\mu$m) and emulsified oil/water ($d<20\mu$m) \cite{chen19}. Depending on the type of mixture different techniques are used in order to separate oil from water \cite{kota12}. 

Gravity separation followed by skimming are typically used to remove stratified oil from water and is considered an efficient, low cost,  primary step in water treatment \cite{cher98, zee83}. For smaller oil droplets these approaches are not effective and follow up steps in treatment are often required. Some of the conventional techniques used to treat emulsions are chemical emulsification \cite{sun98}, centrifugation \cite{cam06,com90}, heat treatment\cite{strom95} and membrane filtration \cite{kota12, yang2019large}. Limitations of these conventional approaches includes high energy costs, operating costs, sludge production and limited efficiency \cite{pa15}.

Recently the role of wettability has been studied in order to propose more efficient and low costs water/oil separation methods. The wetting behavior of a certain surface will depend on the geometry and chemistry of the substrate \cite{Quere2008, we36, we49} and thus, by controlling these key parameters one can develop a material with  antagonistic wetting behavior for oil and water that propitiate the mixture separation.

With that in mind, several materials with special wetting behavior were developed and successfully applied in oil-water separation. This materials can be classified into oil-removing type of materials \cite{feng04,gui10,cortese14,zhan20}, characterized by superhydrophobic/superoleophilic wetting behavior,  and water-removing  type of materials \cite{kota12,yang12}, characterized by superhydrophilic/superoleophobic wetting behavior. Despite the fact that oil-removing materials being more common, they have the disadvantage of been easily fouled by oils due to their oleophilicity nature and not been suitable for gravity-driven separation due to the higher density of the water. On the other hand, water-removing materials are more difficult to achieve due to the fact that most oleophobic materials are also hydrophobic\cite{tuteja07,ahuja08}.

Despite all the advances in this field, most of the studies are focused in the fabrication and performance of these materials and not in the underlying mechanisms that propitiate oil/water separation \cite{chen19}. Therefore, more fundamental research toward understanding the interactions between water, oil and surfaces is extremely necessary in order to build a robust theoretical background that could be used as guideline for further developments in this area.      

In this work we  address this issue studying oil/water separation in two distinct oil-removing surfaces: a pillared surface and a porous substrates.
We concentrate in the case of small volume droplets, for which the dispersed phase size fits into the range of emulsions, where gravity does not play any role in the separation of oil and water. To do that we first apply a theoretical continuous model which takes into account the energy of creating interfaces between solid, liquid and gas phases when a droplet of pure liquid (water or oil) is placed on a substrate. By applying a minimization procedure, we obtain the  wetting state of the droplet that minimizes its energy. This allow us to build a wetting phase diagram for the substrate, which indicates for which range of geometrical parameters the substrate is hydrophobic/oleophilic and the corresponding contact angle of the droplet. To take into account the interaction between water and oil, we simulate a Cellular Potts Model with 4-states using Monte Carlo simulations. It allow us to study the separation capacity of both substrates and evaluate how different geometric parameters affect the performance of these materials in separating oil from water.

This manuscript is organize as follows. In section \ref{theoreticalModel} we present the continuous model and discuss the theoretical results. In section \ref{section_simu} we introduce the Monte Carlo  4-spins cellular Potts model and describe our simulations methods. The simulations results for both surfaces are shown and discussed in section \ref{section_result}.  We close this work with our conclusions and possible extensions of our results in in section \ref{conclusion}.

\section{Theoretical Continuous Model}
\label{theoreticalModel}

The goal of this section is to develop a simple model to determine if a given substrate has the capacity to separate water from oil. An ideal oil-removing material is such that, when a mixed oil/water droplet is deposited on its surface, the oil penetrates the material and the water remains on its surface. It is then important to  identify the wettability states of a substrate.

For this purpose, we first address a following question: if a  droplet of fixed volume $V_0=4/3\pi R_0^3$ and composed by a pure liquid (water or oil) is placed on a rough surface, which is its favorable wetting state? To answer to this question, we assume that there are two possible wetting states, one called Wenzel (W) and characterized by the homogeneous wetting of the surface and one called Cassie-Baxter (CB) with air pockets trapped underneath the droplet. The W state is associated with an omniphilic behavior while the CB state with an omniphobic one.

In this model we  take into account all the interfacial energies associated to the CB and W states and minimize these energies. The wetting state with minimum energy is the one that is favorable from the energetic viewpoint. Similar ideas where used by other authors \cite{Quere2008, Sbragaglia2007, Tsai2010, Shahraz2012} including two of us \cite{fernandes2015, Silvestrini2017, laz19}. 

\begin{figure}[h]
    \includegraphics[width=1\linewidth]{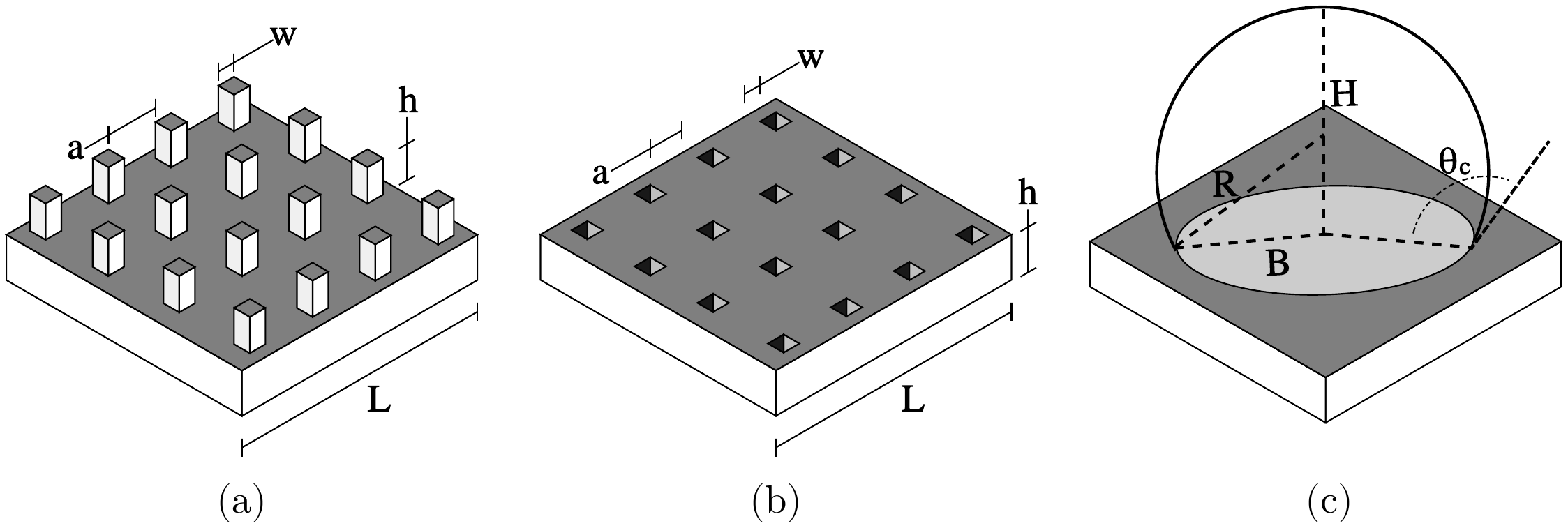}
    \caption{Definition of the geometric parameters of the substrates and the droplet. {\bf (a)}: pillared surface with width $\ww$, pillar distance $\aaa$ and height $\hh$. {\bf (b)}: porous surface with width $\ww$, porous distance $\aaa$ and height $\hh$. {\bf (c)}: geometric parameters of the droplet. We consider a spherical cap with radius $R$, base radius $B$, height $H$, and contact angle $\theta_C$.}
    \label{poros}
\end{figure}

The geometry of the surfaces and the three-dimensional droplet considered in this work are shown in Fig.(\ref{poros}).

The total energy of each of the wetting states is given by the sum of all interfacial energies between the droplet and the desired surface. The energetic difference between the system with and without the droplet is $E^s_{\rm tot} = \Delta E^s + E_g^s$, where superscript $s$ represent the wetting state (CB or W) and $E_g$ is the gravitational energy. $ \Delta E^s$ is the difference in the interfacial energy between every pair formed from liquid, solid, and gas after the droplet is placed on the surface in state $s$ and the energy of the surface without the droplet. The importance of the gravitational energy depends on the droplet size and its composition. In this work we consider droplets with small volumes such that gravitational energy is negligible compared to  $\Delta E^s$.

The interfacial energy equations for pillared substrate, schematized in Fig.(\ref{poros}-a), are developed in Ref.\cite{fernandes2015} and can be written as:

\begin{equation}
\Delta E_{\bf pil}^{\bf {CB}}  = \NCB \left[ (\sSL - \sSG)\ww^2 + (\dd^2-\ww^2)\sGL \right] + \sGL 
\ScapCB
    \label{cb_pilares}
\end{equation}

\begin{equation}
    \Delta E_{\bf pil}^{\bf W} = \NW \left[ (\dd^2+ 4\hh \ww)\sSL \right] + \sGL \ScapW
    \label{w_pilares}
\end{equation}

For the porous surface,  schematized in Fig.(\ref{poros}-b), the interfacial energies are given by:

\begin{equation}
\Delta E_{\bf por}^{\bf {CB}}  = \NCB \left[ (\dd^2-\ww^2)(\sSL - \sSG) + \ww^2\sGL \right] + \sGL 
\ScapCB
    \label{cb_poros}
\end{equation}

\begin{equation}
    \Delta E_{\bf por}^{\bf W} = \NW \left[ (\dd^2 - \ww^2 + 4\hh \ww)(\sSL - \sSG) + \ww ^2 \sGL \right] + \sGL \ScapW
        \label{w_poros}
\end{equation}

For all equations above, $\dd = \ww+\aaa$  and $\sSG$, $\sSL$ and $\sGL$ are the surface tensions for the solid-gas, solid-liquid and liquid-gas interfaces, respectively.  The subscript "L" accounts for a liquid phase and can be either water of oil. $\Ns=\frac{\pi B^2}{d^2}$ accounts for the number of pillars/pores underneath the droplet and $\Scap=2\pi R^2[1-\cos(\theta^s_c)]$ is the surface area of the spherical cap, where $B^s=R^s \sin(\theta^s_c)$ is the base radius and $\theta^s_c$ is the contact angle of the droplet, as defined in Fig.(\ref{poros}-c).  

In order to identify the favorable wetting state from the thermodynamic point of view for the pillared surface, we minimize Eqs.(\ref{cb_pilares}, \ref{w_pilares}), and compare the global minimal energy for each state. This energy minimization process is explained in ref. \cite{fernandes2015,Silvestrini2017} and in the Supplementary material. For the porous substrate, the process is analogous but we use Eqs.(\ref{cb_poros}, \ref{w_poros}). Besides predicting the favorable wetting state, this approach also allows the determination of the contact angle, $\theta_c$, associated to the most stable state. 

The continuous model and the minimization process are employed to build a theoretical wetting diagram for both substrates considered in this work. 
For the calculations we considered surface tensions obtained from experiments with water droplets on poly-(dimethylsiloxane) (PDMS) surface \cite{Tsai2010} and with hexadecane oil droplets on the same PDMS surface\cite{martin17}. Solid-liquid surface tensions were obtained from Young?s relation $\sGL \cos(\theta) = \sSG - \sSL$, where $\theta$ is the contact angle on a smooth surface and assumes the values $\theta^w = 114^{\circ}$ for water\cite{Tsai2010} and $\theta^o = 53^{\circ}$ for oil\cite{martin17}. Water/hexadecane surface tension were obtained from Ref. \cite{wu99}. These surface tensions are summarized in table \ref{surface_tention}. We fix $\ww=5\mu m$ and screen over the parameters $\aaa \in (0,16]\mu m$ and $\hh \in (0,15]\mu m$.

\begin{table}
    \centering
    \begin{tabular}{|c||c|c|c|}
    \hline
         &   Water   & Solid    & Gas \\
    \hline
    \hline
    Oil  & $\sWO=$ 53.5 &  $\sSO=$ 8.6 & $\sGO=$27  \\
  Water  &  ---        & $\sSW=$50.2 & $\sGW=$70    \\
   Solid & ---   & ---       & $\sSG=$ 25  \\
    \hline
    \end{tabular}
    \caption{Surface tensions used in this work in units of nN/m.}
    \label{surface_tention}
\end{table}

\begin{figure}[h]
\centering
\includegraphics[width=1\linewidth]{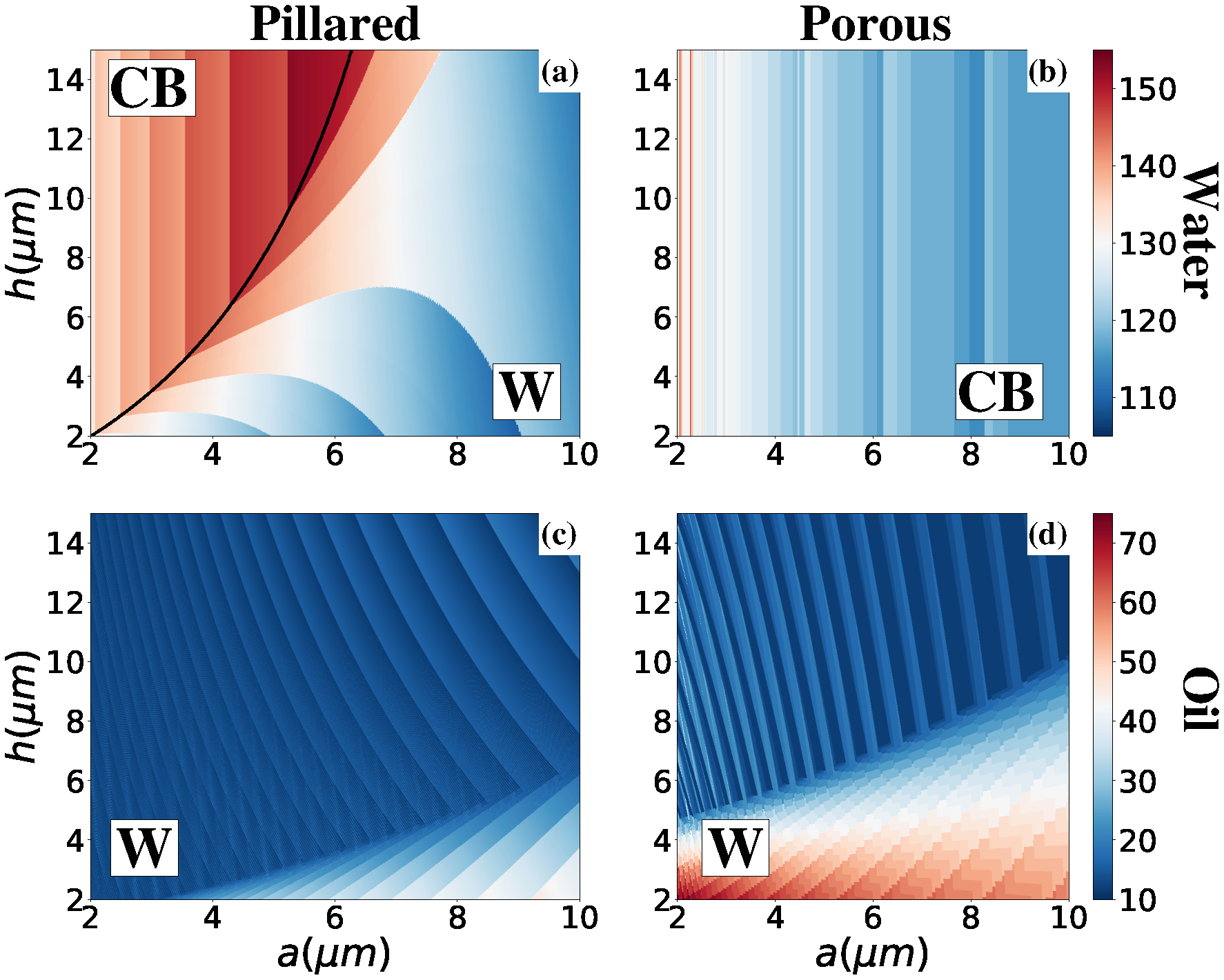}
\caption{Theoretical wetting diagrams for {\bf (a)} water droplet placed on a pillared surface, {\bf (b)} water droplet placed on a porous surface, {\bf (c)} oil droplet placed on a pillared surface and {\bf (d)} oil droplet placed on a porous surface, as a function of two geometrical parameters of the surface: the height of the pillars $h$, and the interpillar/interpourous distance $a$. Pillars width are kept constant $\ww=5\mu m$. The solid line in {\bf (a)} represents the predicted thermodynamic transition between the Cassie-Baxter (CB) and  (W) states. Colors indicate the droplet's contact angle, in degrees. }
\label{teorico_pillar}
\end{figure}

Fig.(\ref{teorico_pillar}) summarizes our theoretical results for pure water/oil droplets with $R_o=50 \mu$m. We first concentrate in the case of the pillared surface and consider a droplet with pure water. In this case, the surface presents two regions as shown in Fig.(\ref{teorico_pillar}-a): for small $\aaa$ and high $\hh$, the  favorable state is CB, but, as the pillar distance $\aaa$ increases a transition to W state is observed. This transition depends on the initial size of the droplet as shown in Ref. \cite{fernandes2015}. When the droplet is composed by pure oil, W is the  favorable state for any geometric parameter, as shown in  Fig.(\ref{teorico_pillar}-c).

The same process is applied for the porous substrate. When we consider a droplet of pure water no phase transition is observed and the favorable wetting state is CB for any geometric parameter as shown in Fig.(\ref{teorico_pillar}-b). Nevertheless lower contact angles were predicted when compared to the pillared surface indicating that this surface is less hydrophobic. If a droplet of pure oil is taken into account, the  thermodynamic state is W in the whole diagram (Fig.\ref{teorico_pillar}-d). In comparison with the pillared surface, higher values of the contact angle were observed for low values of porous height $\hh$.

The analysis presented here indicates that the porous surface could function as a good oil removing material regardless the choice of the surface geometrical parameters, but higher values of porous height $\hh$ would favors the separation due to the lower contact angles predicted for the oil droplets. For the pillared surface we expect a good oil/water separation in the region marked as CB in Fig.(\ref{teorico_pillar}-a). 

However, this approach has the limitation of only considering  pure water or pure oil droplets,  disregarding the effects of water-oil interaction.In order to overcome this limitation we perform numerical simulations of the cellular Potts model taking into account the promising interval of parameters obtained by the theoretical analysis discussed in this section. 

\section{Simulations: 4-spins Cellular Potts model}
\label{section_simu}

Monte Carlo simulations (MC) of the cellular Potts model have been used to study wetting phenomena in textured surface \cite{Lopes2013,Oliveira2011,Mortazavi2013} but is also a useful tool for example to study cell migration on substrates  \cite{Graner1992,fortuna2020,magno2015}. The coarse-grained approach used in this types of simulations (in opposition to the explicit atoms approach commonly used in molecular dynamic simulations) is a more consistent framework to treat mesoscopic systems and, therefore, more appropriate for comparison with experimental results.


Here we expand the  2D Potts model developed by Mombach {\it et al.} \cite{Lopes2013,Oliveira2011} and the 3D cellular Potts model proposed to simulate a pure liquid droplet in a specific surface\cite{fernandes2015, Silvestrini2017} in order to treat a droplet composed of a mixture of two liquids: water and oil. Our model consist in  a simple cubic lattice in which each state represents one of the components: gas, water, oil, or solid. The Hamiltonian used is

\begin{eqnarray}
H &=& \frac{1}{2} \sum_{<{\rm i},{\rm j}>} E_{s_{\rm i},s_{\rm j}}(1-\delta_{s_{\rm i},s_{\rm j}}) + \alpha_w \left( \sum_{\rm i} \delta_{s_{\rm i},1}-V^w_T \right) ^2  \nonumber \\
&+& \alpha_o \left( \sum_{\rm i} \delta_{s_{\rm i},2}-V^o_T \right) ^2 + g \sum_{\rm i} (m_{\rm i} h_{\rm i} \delta_{s_i,1}  + m_{\rm i} h_{\rm i} \delta_{s_{\rm i},2} )
\label{hamil}
\end{eqnarray}

\noindent where the spin $s_i \in \{0,1,2,9\}$ represent gas, water, oil and solid states, respectively.

The first term in Eq.(\ref{hamil}) represents the energy related to the presence of interfaces between sites of different types. The summation ranges over pairs of neighbors which comprise the 3D Moore neighborhood in the simple cubic lattice (26 sites, excluding the central one), $E_{s_i,s_j}$ is the interaction energies of sites $s_i$ and $s_j$ of different states at interfaces and  $\delta_{s_i,s_j}$ is the Kronecker delta. 

In second and third terms in Eq.(\ref{hamil}), $V^w_T$ and $V^o_T$  are the target water and oil volumes, respectively, the summations are the water and oil volume and the parameters $\alpha_w$  and $\alpha_o$ mimics the liquids compressibility.  Thus, these terms maintain the liquids volumes and the desired composition of the droplet constant during the simulation. The last term  is the gravitational energy, where $g = 10m/s^2$  is the acceleration of gravity and $m_i$ is the mass of the site. In both the volumetric and gravitational terms, only sites with liquid, $s_i = 1$ or $s_i = 2$, contribute.

In our simulations the length scale is such that one lattice spacing corresponds to 1 $\mu$m and the surface tensions values (shown in Table \ref{surface_tention}) are divided by 26, which is the number of neighborhoods that contributes to the first summation of our Hamiltonian. Therefore, the interfacial interaction energies $E_{s_i,s_j} = A\sigma_{s_is_j}$, with $A=1\mu$m$^2$ are given by $E_{0,1} = 2.70 \times 10^{-9}\mu$J, $E_{0,2} = 1.04 \times 10^{-9}\mu$J, $E_{0,9} = 0.96 \times 10^{-9}\mu$J, $E_{1,2} = 2.06\times 10^{-9}\mu$J, $E_{1,9} = 1.93\times 10^{-9}\mu$J and $E_{2,9} = 0.33\times 10^{-9}\mu$J. The mass existent in a unit cube is $m^w = 10^{-15}$kg for water and  $m^o = 0.77 \times 10^{-15}$kg for oil. We fix $\alpha_w = \alpha_o = 0.01 \times 10^{-9}\mu$J/($\mu$m$)^6$ (the choice of these values are justified in the supplementary material).

The total run of a simulation is $5 \times 10^{5}$ Monte Carlo steps (MCS), from which the last $2.5 \times 10^{5}$ MCSs are used to measure observables of interest. Each MCS is composed by $V^T = V^w + V^o$ number of trial spin flips, where $V^T$ is the volume of the liquid droplet which is composed by a volume of oil $V^o$ and of water $V^w$. A spin flip is accepted with probability $ \text{min}\{1,\text{exp}(- \beta \Delta H)\} $, where $\beta = 1/T$. In the cellular Potts model $T$ acts as noise to allow the phase space to be explored. In our simulations, a value of $T=9$ was used, which allows an acceptance rate of approximately 15$\%$ while keeping both water and oil in a liquid state (for further information see supplementary material). 

\begin{figure}
\centering
    \includegraphics[width=1\linewidth]{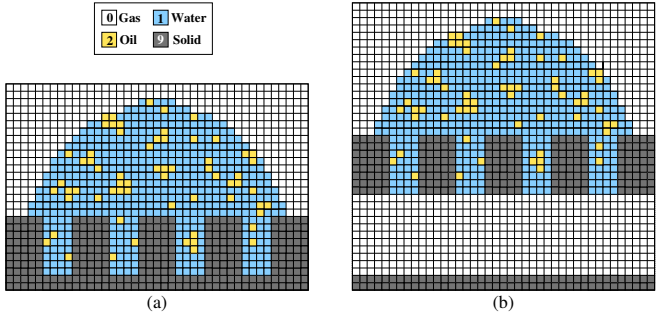}
    \caption{Visual Schematic of the initial set up of the simulations for {\bf (a)} the pillared surface and {\bf (b)} the porous surface.}
    \label{CI}
\end{figure}

The initial wetting state is created using a hemisphere with initial volume $V^T  \approx V_0=4/3\,\pi R_0^3$, due to the discreteness of the lattice. The droplet has $R_0=50 \mu$m  in a cubic box with $L=240\mu m$. The composition of the droplet is defined by the oil fraction $f_o$, thus, $V^o = f_o V^T$ is the oil volume. One can also define the water volume  $V^w = f_w V^T$, where $f_w=1-f_o$ is the water fraction. Oil and water sites are randomly distributed in the droplet. As in the previous section, two different substrates are studied as possible oil-removing materials: pillared surface and porous surface. When the porous surface is used a reservoir with $V_{\rm res}>3V^T$ is added to the  bottom of the surface, as shown in Fig.(\ref{CI}-b).

\subsection{Definitions of the efficiencies}

In order to evaluate the efficiency of the different substrates, we calculated the percentage of the initial oil/water volume that is between the pillars or inside pores, $V^l_p$, and, for the porous case, the percentage of the initial oil/water volume that is inside the reservoir, $V^l_r$.  The superscription $l$ refers to water, $w$, or oil, $o$. These quantities are calculated as follow:

\begin{equation}
V^l_p = n^l_p/V^l
\end{equation}
\begin{equation}
V^l_r = n^l_r/V^l
\end{equation}

\noindent where $n^l_p$ is the number of liquid sites in between the pillar or inside the pores and $n^l_r$ is the number of liquid sites inside the reservoir. This quantities allows us two define a liquid absorption efficiency for the pillared and porous surface:

\begin{equation}
\epsilon^l_{{\bf pil}} = V^l_p
\end{equation}

\begin{equation}
\epsilon^l_{{\bf por}} = V^l_p + V^l_r
\end{equation}

The ideal substrate for oil and water separation , in our case, is such that all the water sites remain above the surface while all the oil sites are adsorbed by the substrate  and so, we propose that the separation efficiency can by written as:

\begin{equation}
\xi_{{\bf S}} = \frac{\epsilon^o_{{\bf S}} + (1-\epsilon^w_{{\bf S}})}{2}
\label{eff_simu}
\end{equation}

\noindent where the index $S$ refers to the pillared or porous surface.

We also calculate in our simulations two efficiencies that are commonly used in experiments. The first one measures the amount of water that is not absorbed by the substrate \cite{gu2014robust,singh2016fabrication}. In our case, this efficiency is calculated as follow:

\begin{equation}
\xi^{\bf a}_{\bf pil} = 1 - V_p^w
\label{eff_abv_pil}
\end{equation}

\begin{equation}
\xi^{\bf a}_{\bf por} = 1 - V_p^w - V_r^w
\label{eff_abv_por}
\end{equation}

The second one measures the oil rejection coefficient given by $R=(1-C_p/C_o)$, where $C_o$ is the initial concentration of oil, which in our case is $f_o$, and $C_p$ is the concentration of oil in the remaining water above the surface\cite{wang2017novel,su2019rubber}. In our simulation  this efficiency is calculated as follow:

\begin{equation}
\xi^{\bf r}_{\bf pil} = 1 - \left[\frac{V^o - n_p^o}{ V^T-[n_p^w+n_p^o]}\frac{1}{f_o}\right]
\label{eff_rej_pil}
\end{equation}

\begin{equation}
\xi^{\bf r}_{\bf por} = 1 - \left[\frac{V^o  - (n_p^o + n_r^o)}{ V^T-[(n_p^w+n_r^w)+(n_p^o+n_r^o)]}\frac{1}{f_o}\right]
\label{eff_rej_por}
\end{equation}

\section{Results and Discussion}
\label{section_result}

In this section we analyze the simulation results for pillared surfaces, Fig.(\ref{poros}-a), and porous surfaces, exemplified in Fig.(\ref{poros}-b). We compare these results with the theoretical predictions presented in the section \ref{theoreticalModel} and discuss the efficiency of these two types of surfaces in separating water from oil.

\subsection{Pillared Surface}

The theoretical results summarized in Fig.(\ref{teorico_pillar}) show that, if a pure water droplet is placed on a pillared surface, it presents a CB and  W regions depending on the substrate's geometrical parameters. As previously explained, these wetting stares are associated with a hydrophilic and hydrophobic behaviors respectively. For any geometric parameter, if a pure oil droplet is deposited on the substrate, its favorable state is W, indicating that these substrates are oleophilic. Then, we expect that pillared surfaces could  work as an oil removing material in the region where a pure water droplet is in a CB state. If this is the case, these substrates would act as "sponges", absorbing oil and leaving water above the pillars.

\begin{figure*}
\centering
\includegraphics[width=0.9\linewidth]{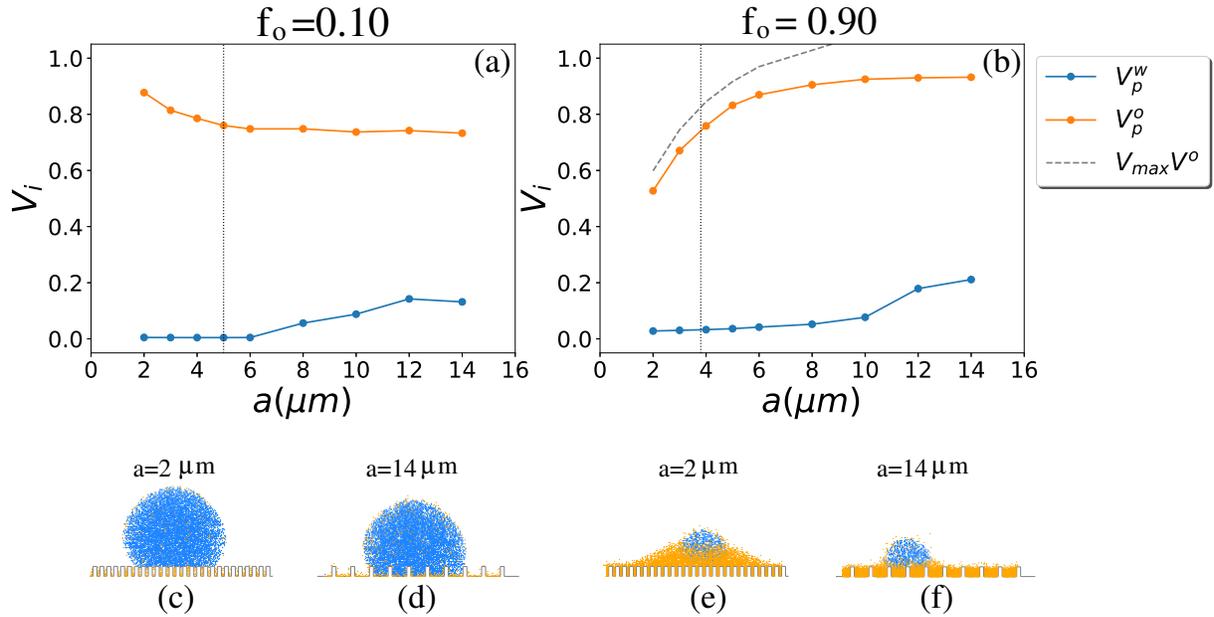}
\caption{{\bf Results for pillared substrates for several geometric parameters.} Above: Interpillar volume of water, V$^w_p$, and oil, V$^o_p$, as a function of the pillar distance $\aaa$ for (a) $f_o = 0.10$ and (b) $f_o = 0.90$. Below each figure: Cross sections of the final droplet configuration of Monte Carlo simulations for $a=2 \mu$m and  $a=14 \mu$m for each corresponding $f_o$. To build this figure, a mixed droplet with oil fraction given by $f_o$  (value specified above the  figures) and total volume given by $V^T=4/3\pi R_0^3$ with $R_0 = 50 \mu$m is simulated on a substrate with $\ww=5 \mu$m,  $\hh=10 \mu$m and varying pillar distance $\aaa$. Blue color represent water and orange represent oil.}
\label{res_pillar}
\end{figure*}

With that in mind, here we study surfaces with fixed pillar height $\hh=10 \mu$m and pillar width $\ww=5 \mu$m and  several values of  pillar distance $\aaa$. In the SM we also show the analyses for the case with $\hh=5 \mu$m, which presents very similar results. Fig.(\ref{res_pillar})  shows the interpillar volume, V$^l_p$ as a function of the pillar distance $\aaa$ for $f_o = 0.10$ (Fig.\ref{res_pillar}-a) and $f_o =0.90$ (Fig.\ref{res_pillar}-b) and cross sections of the droplet configuration in the final state of the Monte Carlo simulations correspondent to two different pillar distance, $\aaa=2\mu m$ and $\aaa=14\mu m$, for each $f_o$. The vertical gray dotted lines shown in Fig.(\ref{res_pillar}-a,b) indicate the water CB-W transition predicted by the theoretical model  for the correspondent water volume.


The available volume to absorb oil is the maximum volume between pillars, given by $V_{\bf max}=(\frac{L}{\dd})^2(\dd^2-\ww^2)\hh$.  We normalize it by the total oil volume present in the droplet $V^o$, $V_{\bf max}/V^o$, and show this quantity in Fig.(\ref{res_pillar}-b), represented by the gray dashed line. This curve does not appear for the smaller oil fraction $f_o$ because $V_{\bf max} \gg V^o$ for all values of $\aaa$. 

For both cases we observe that for low values of interpillar distance $\aaa$ water does not penetrates the pore, V$^w_p \approx 0$,  (Figs.\ref{res_pillar}-a,b), which is consistent with a CB wetting state. For higher values of the interpillar distance $\aaa$ we observed an increase in V$^w_p$, that roughly coincides with the theoretical prediction from CB to W states indicated by the vertical line in Figs.\ref{res_pillar}-a,b. This is visually confirmed by the cross sections of the final configurations shown in Figs.(\ref{res_pillar}-c,d) where the water is in the CB state for $\aaa=2\mu m$ and in the W state for $\aaa=14\mu m$. We have checked that it also happens for  $f_o=0.9$, Figs.(\ref{res_pillar}-e,f), but it is not possible to visualize because the oil sites dominates the image and do not allow us to properly see the water behavior. 

Although we obtain a nice agreement with the theoretical results for the water behavior, we note that, when considering the mixture, the phase transition occurs for higher values of the interpillar distance $\aaa$. This discrepancy is expected because the wetting diagram is built for a pure water droplet or pure oil droplet, while in the simulations there is a composition of both liquids. 

We now discuss the oil behavior. Theoretical calculations show that a pure oil droplet does not undergo any wetting state transition, remaining in the W state for all values of interpillar distance $\aaa$. This is qualitatively confirmed by simulations, as shown by the oil penetration in Figs.(\ref{res_pillar}-a,b). For $f_o=0.10$ (Fig.\ref{res_pillar}-a) and low values of the interpillar distance $\aaa$,  V$^o_p$ indicates that 88\% of the initial oil volume penetrates the substrate. As $\aaa$ is increased, the percentage decreases and a plateau is observed at V$^o_p \approx 0.73$. For the case with $f_o=0.9$ we observe an increase of V$^o_p$ with the increase of $\aaa$ and a plateau is reached at V$^o_p \approx 0.93$. This is related to the fact that the available volume for oil absorption, $V_{\bf max}$, is of the same order of the total amount of oil in the droplet, as indicated by the dashed line in Fig.(\ref{res_pillar}-b) and so, the substrate becomes saturated.

Despite of the high percentage of the oil absorbed by the surface, we note that from 7\% to 27\% of the initial oil remains above the surface. We remark that there is no oil inside the droplet of water: all the remaining oil is at the interface of the droplet, which creates an water-oil-gas interface. We note that reminiscent oil in the remaining water was also observed experimentally\cite{gondal2014study,liu2016facile}

To understand this feature in our simulations,  we analysed the terms of Eq.(\ref{hamil}) related to the energy for creating interfaces and evaluated the necessary conditions for the appearance of a spin of type "oil" on the interface between the water and the gas. We first consider a spin  $\rm i$ with type gas G that is on the interface of the droplet and has  $\NvizO$ neighbors of type "oil", $\NvizG$  neighbors of type "gas" and $\NvizW$  neighbors of type "water". The flip of this spin $\rm i$ from a gas to oil state is energetically favoured when: 

\begin{equation}
    \NvizO > \NvizG-\NvizW\frac{(\sGW-\sWO)}{\sGO}. 
\end{equation}
 
This lead to two conclusions: (i) The presence of the oil is favorable favorable when the site $\rm i$ is surrounded by other spins of type oil.  It shows that for the oil on the interface is favorable to form a cluster or a film. (ii) Even if the site $\rm i$ has no oil neighbor, $\NvizO=0$,  the presence of the oil is energetically favorable when  $\NvizG < \NvizW((\sGW-\sWO)/\sGO)$. This last condition is mathematically possible when $\sGW > \sWO$, which is the physical case as can be seen in Table \ref{surface_tention}. Physically, it tells us that the appearance of a oil on the interface can happen because  $\sGW$ is  high compared to $\sWO$. This suggests that changing the gas in such way to increase $\sWO$ and/or decrease $\sGW$ could improve the capacity of separating oil and water. 
 
A very similar calculation can be done for the case where the spin $\rm i$ is water and we compute the conditions at which it is energetically favorable to change to a spin of type oil: $\NvizO > \NvizW - \NvizG ((\sGW-\sGO)/\sWO)$.  The analysis of this equation leads to the same conclusions described above.

To end this section, we discuss the efficiency of this type of substrate using three different definitions. We measure $\xi^{\bf a}_{\bf pil}$ defined in Eq(\ref{eff_abv_por}), which measures the amount of water that is not absorbed by the surface and $\xi^{\bf r}_{\bf pil}$ that measures the capacity of the surface to exclude oil from the water that remains above the surface. We compare these quantities with the proposed separation efficiency given by the Eq.(\ref{eff_simu}),  which takes into account both the capacity of maintaining water above the substrate and the capacity of absorbing oil. 

\begin{figure}
\centering
\includegraphics[width=1.0\linewidth]{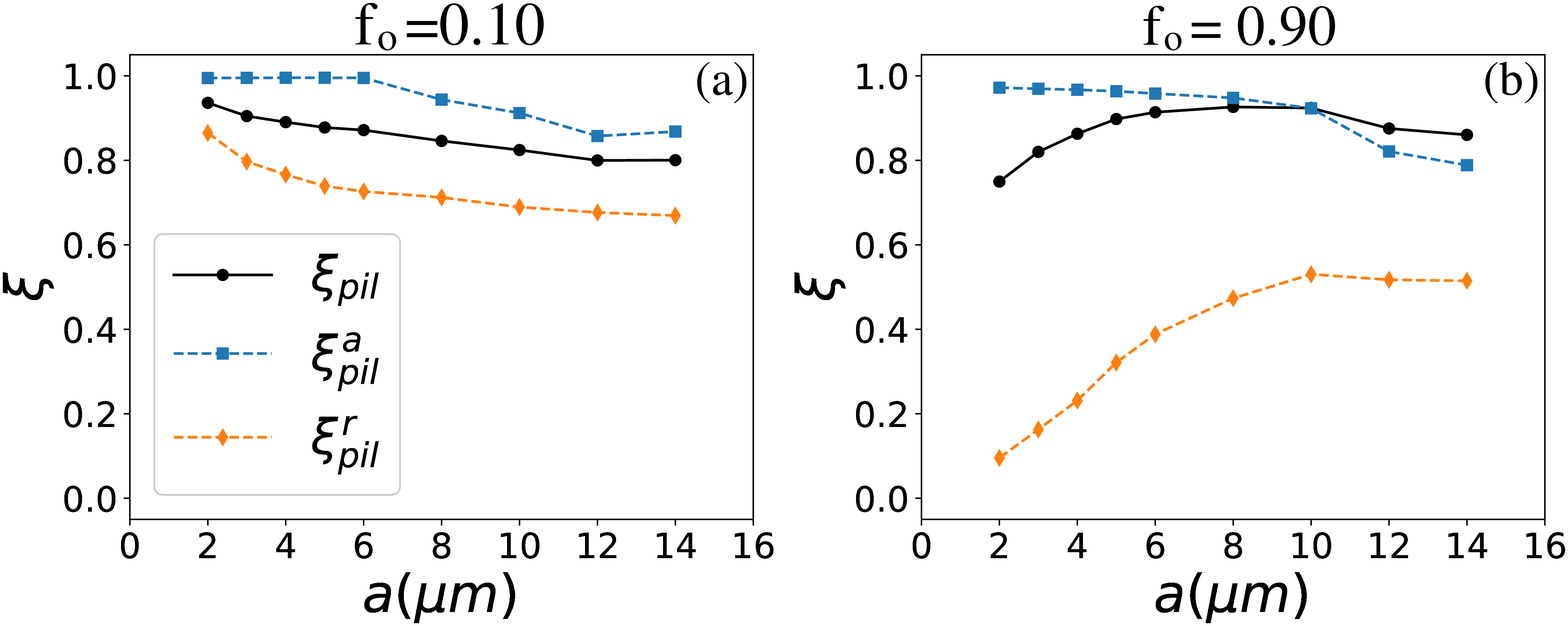}
\caption{{\bf Efficiency  for pillared substrates.} $\xi^{\bf a}_{\bf pil}$, $\xi^{\bf r}_{\bf pil}$ and  $\xi_{\bf pil}$ as a function of interpillar distance $\aaa$ for (a)$f_o=0.10$ and (b)$f_o=0.90$.}
\label{eff_pillar}
\end{figure}

Fig.(\ref{eff_pillar}) summarizes our results for all geometries considered here. For both values of $f_o$, $\xi^{\bf a}_{\bf pil}$ follows the behavior of water: it shows high efficiency when the water is in the CB state and decays when the transition to W occurs. The $\xi^{\bf r}_{\bf pil}$, on the other hand, follows the behavior of the oil where the efficiency is high for surfaces where V$^o_p$ is also high.

These measurements have the disadvantage of only considering one aspect of the separation process, which can be misleading. For instance, for the case of $f_o=0.90$ and $\aaa=2 \mu$m  if we consider $\xi^{\bf a}_{\bf pil}$ the efficiency of this substrate is approximately 100\% despite of the amount of oil that remains above the surface (Fig.(\ref{res_pillar}-e)) because there is no absorption of water. On the other hand,  if we consider $\xi^{\bf r}_{\bf pil}$ the efficiency would be very low even though the pillared surface is able to absorb 53\% of the initial oil volume and repel almost all the water.

The efficiency $\xi_{\bf pil}$ proposed in this work for oil removing materials takes into account all mechanisms that contributes to the water/oil separation. Thus, considering the same case of $f_o=0.90$ and $\aaa=2 \mu$m discussed above,  $\xi_{\bf pil}$ is lower than $\xi^{\bf a}_{\bf pil}$ because it considers the reminiscent oil above the surface and  $\xi_{\bf pil}$ is greater than $\xi^{\bf r}_{\bf pil}$  because it considers the quantity and purity of the absorbed oil.

Despite of the good efficiency observed for some these pillared surface, they have the limitation of only been able to absorb a certain volume of oil,  $V_{\bf max}$. In the next section we evaluate the performance of a surface which in principle do not have this problem.

\subsection{Porous Surface}

\begin{figure*}
\centering
\includegraphics[width=0.9\linewidth]{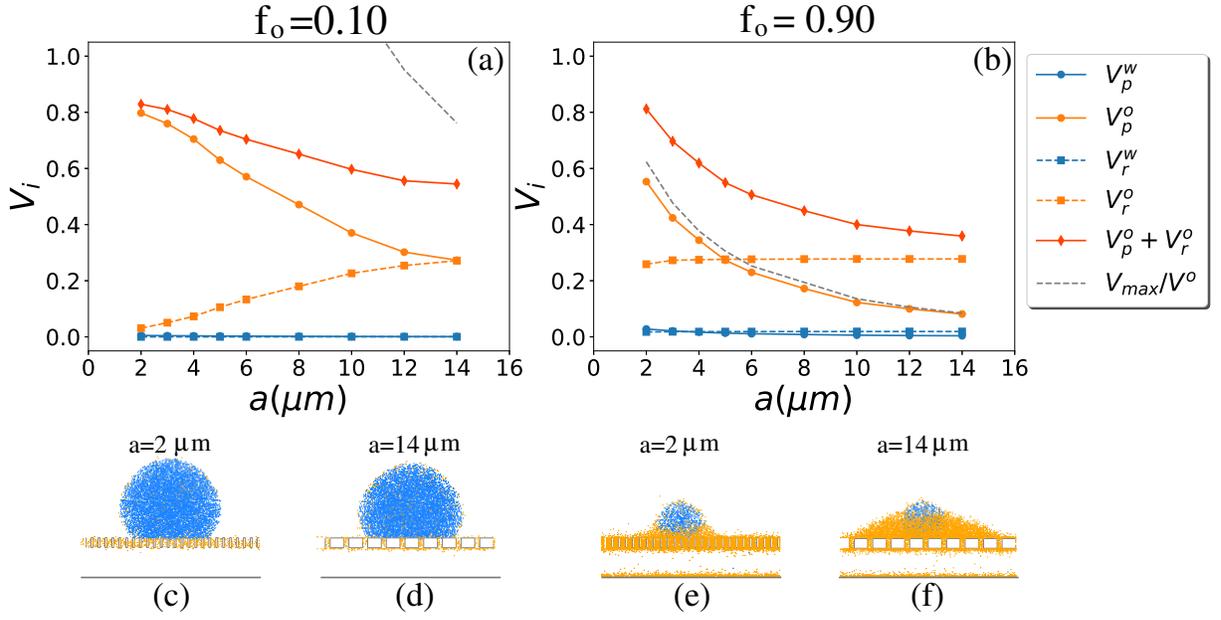}
\caption{{\bf Results for porous substrates for several geometric parameters.} Above: Interporous volume of water V$^w_p$  and oil V$^o_p$ and reservoir volume of water V$^w_r$  and oil  V$^o_r$  as a function of the pillar distance $\aaa$ for (a) $f_o = 0.10$ and (b) $f_o = 0.90$. Below each figure: Cross sections of the final droplet configuration of Monte Carlo simulations for $a=2 \mu$m and  $a=14 \mu$m for each corresponding $f_o$. To build this figure, a mixed droplet with oil fraction given by $f_o$  (value specified above the  figures) and volume given by $V^T=4/3\pi R_0^3$ with $R_0 = 50 \mu$m is simulated on a substrate with $\ww=5 \mu$m,  $\hh=10 \mu$m and varying porous distance $\aaa$. Blue color represent water and orange represent oil.}
\label{v_poro}
\end{figure*}

In this section we consider a porous substrate where the oil can be drained into a reservoir. Here we explore this surface for the same oil fractions $f_o$  and similar geometrical parameters considered for the pillared surface: $\ww=5 \mu$m, $\hh=10 \mu$m and several values of porous distance $\aaa$. According to the theoretical predictions, for pure water or pure oil, there is no wetting transition: the porous surface is hydrophobic and oleophilic for all values of geometric parameters, as summarized  in Figs.(\ref{teorico_pillar}-b,d).

Fig.(\ref{v_poro}) shows the interporous volume V$^l_p$ and the reservoir volume V$^l_r$ as a function of porous distance $\aaa$ for two oil fractions, $f_o=0.10$ and $0.90$. For this type of surface,  the volume normalization is such that  V$^l_p$ + V$^l_r$  + V$^l_{ab} =1$ where V$^l_{ab}$ is the remaining volume above the surface. 
For  water, we observe that V$^w_p \approx 0$ and  V$^w_r \approx 0$ which means that water remains above the surface and do not enter the porous or the reservoir for both values of $f_o$ and all values of $\aaa$. This agrees with the theoretical predictions for pure water on porous substrates shown in Fig.(\ref{teorico_pillar}-b), which indicates they are hydrophobic for all geometric parameters. 

Concerning the oil behavior, Fig.(\ref{v_poro}) shows its presence in the pores  V$^o_p$ and in the reservoir V$^o_r$ separately and also the sum of both contributions. This figure also presents the volume of the pores (dashed line). Two things dictates the oil behavior for the porous surface: (i) the available volume inside the pores and (ii) the solid surface area above and bellow the substrate. For small values of the porous distance $\aaa$ we have more interporous volume available for the oil and low solid surface and thus, the oil remains inside the pores. As we increase $\aaa$, V$^o_p$ decreases due to the limited volume of the porous and V$^o_r$ increases due to the increase of the solid surface. However, the increase of the solid surface allows for a formation of an oil film on the surface as well which jeopardize the entry of the oil in the reservoir. Since gravity does not play any role for this volume size, once the porous are filled with oil, it creates a layer that prevent the rest of oil present above the substrate to be absorbed and stored in the reservoir. We stress that there is also a part of the oil which remains as a film in the interface between water and gas as  discussed in the previous section.

\begin{figure}
\centering
\includegraphics[width=1\linewidth]{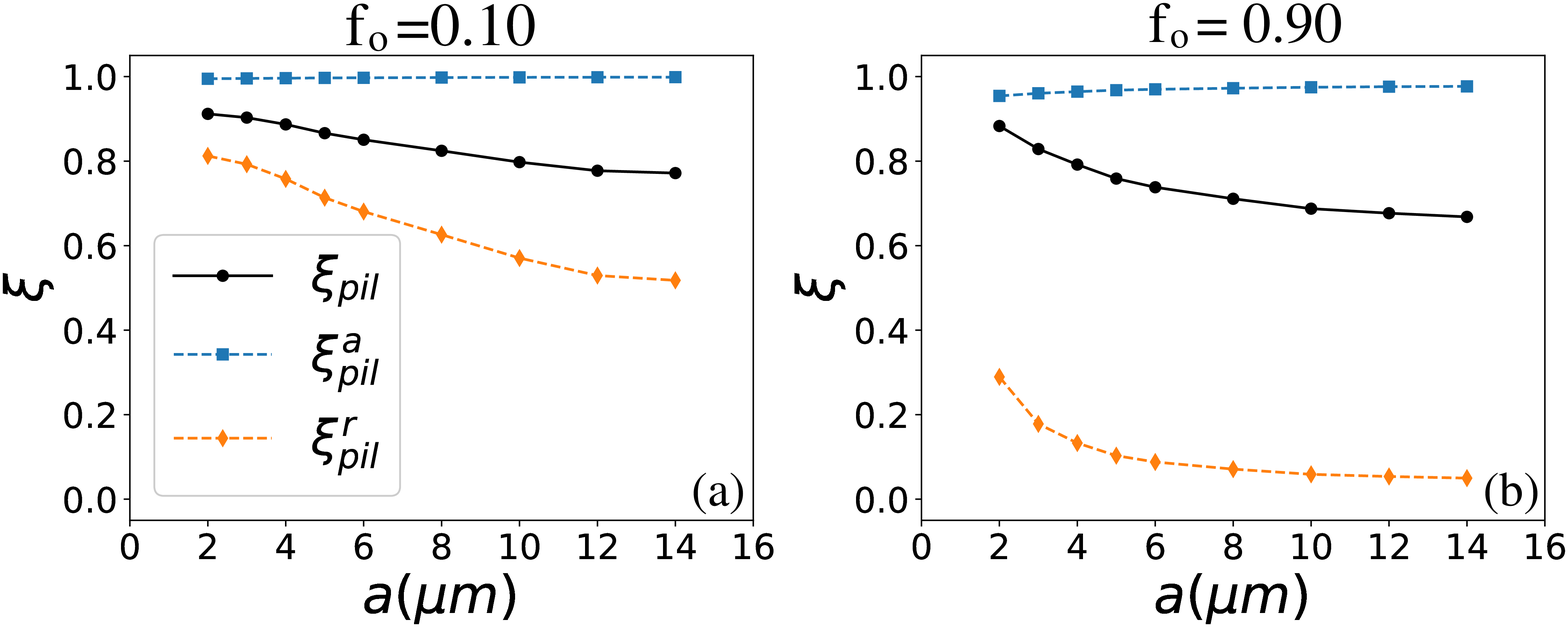}
\caption{{\bf Results for porous substrates.} $\xi^{\bf a}_{\bf por}$, $\xi^{\bf r}_{\bf por}$ and  $\xi_{\bf por}$ as a function of interpillar distance $\aaa$ for (a)$f_o=0.10$ and (b)$f_o=0.90$  }
\label{eff_porous}
\end{figure}

Fig.(\ref{eff_porous}) shows the efficiency of these surfaces in terms of the three measures defined previously. Similarly to the pillared case, $\xi^{\bf a}_{\bf por}$ and $\xi^{\bf r}_{\bf por}$ follows the behavior of the water and oil absorption respectively. In other words, these quantities only reflects the hydrophobicity or oilphilicity of the substrate and, therefore, ignore part of the mechanisms involve in water/oil separation. The alternative definition $\xi_{\bf por}$ takes into account both the hydrophobicity and oilphilicity by considering the  total of oil present in the droplet to define an efficiency, allowing to have an idea of the reminiscent oil above besides the information about the amount of water below the substrate. 

\section{Conclusions}
\label{conclusion}

In this work we propose a theoretical  and a numerical model to investigate the efficiency of a substrate in separating oil from water by controlling its wetting properties. We apply these methods in pillared and porous substrates. We first used the theoretical model \cite{fernandes2015, Silvestrini2017} to investigate the wetting behavior of the substrate when a droplet of oil or water is deposited on it. This approach takes into account the energy of creating surfaces between the liquid, the gas and the solid and employ a minimization procedure to obtain the thermodynamic wetting state of a droplet, together with a determination of its contact angle. With this method, we built a wetting state diagram of the substrate, which indicates the range of geometric parameters for which the  substrate is hydrophobic or oleophilic. With the idea that an appropriate material to separate oil from water would be oleophilic and hydrophobic, this diagram  guides us in choosing the right type of substrates with the correct range of parameters. Because the theoretical approach does not take into account the interaction between water and oil,  we implemented Monte Carlo simulations of a Potts model in three dimensions. The simulations allow us to study the efficiency of the substrates in separating water from oil at different fractions of both liquids.

The theoretical analyses show that pillared substrates present a hydrophobic and a hydrophilic phase (Fig.\ref{teorico_pillar}-a), and are oleophilic for all geometric parameters (Fig.\ref{teorico_pillar}-c). Porous substrate are hydrophobic and oleophilic for all geometric parameters  (Fig.\ref{teorico_pillar}-a,c). We then simulated a droplet with fixed volume size but different oil fractions on both types of substrates. In this work we concentrate in the case where the droplet is small enough to guarantee that gravity is not relevant compared to the energy scales of the interfacial energies. Overall, our simulations results shown in Fig.(\ref{res_pillar}) and Fig.(\ref{v_poro})  and compared to the wetting diagrams, allow us to conclude that the water behavior can be explained by the hydrophobicity of the substrate but the oil behavior is more complex than just evaluating if the substrate is oleophilic. In other words, if the substrate is hydrophobic when tested with pure water, when a mixed droplet composed by water and oil is deposited on it, water remains above the substrate. Predicting the oil behavior is more difficult because the absorption capacity  of the surfaces are limited for two reasons: i) some part the oil remains on the droplet, forming a film between the water and the gas phase and ii) in absence of gravity, once the substrate is filled with the oil, even in the presence of a reservoir (which is the case of our porous substrate), the oil does not fill it completely.

Concerning the efficiency of these substrates in separating water from oil, there are some ways to define it  and the  goodness of the definition depends on what is aimed to capture. Here we explore two different ways that this efficiency is measured in experiments: one that measures the percentage of water that isn't absorbed by the substrate, $\xi_{a}$, and one that measures the capacity of the material to remove oil from the remaining water above the surface, $\xi_{r}$. This measurements have the disadvantage of only explore one aspect that favors the oil/water separation. To capture both relevant oil/water separation mechanisms we proposed a different definition of efficiency called separation efficiency, $\xi_{}$. This observation aims to call the attention to the fact that efficiency very close to 1 does not guarantees that all the important separations mechanisms are been accounted for. 

Both theoretical and simulations models can be modified to analyse other types of substrates and to take into account some effects that are not considered in this work. We discuss some examples and their interests in the following. We studied in this work situations where the mass of oil is small enough for gravity not being relevant, which is the case for emulsions for example.  For these situations, a superhydrophobic fractal substrate\cite{Kao1995, fractalgrowthlivro1995} could act as an  efficient sponge to absorb oil from water, because it maximizes the contact surface between oil and the solid. We could also analyse the situation where the volume of oil is bigger and gravity would be an important element. This can be introduced in our model by changing the length scale for the sites and it would allow  to access phenomena that happens in free oil/water mixtures and dispersion. Another possible adaptation of this model is the study of underwater wetting phenomena, which can be of practical interest if the separation of oil and water happens without a presence of gas \cite{kota12, chen19}.

\section*{Data Availability Statement}

The data that supports the findings of this study are available within the article and its supplementary material. 

\section*{Dedication}

This paper is dedicated to Professor Marcia Cristina Bernardes Barbosa, an outstanding  Brazilian physicist who, not only has greatly contributed to our knowledge of the water behavior, but also actively acts against gender inequality in the field. She inspires many women, including the authors of this manuscript, to pursue a scientific career.

\begin{acknowledgments}
We thank the Brazilian agency CAPES and CNPq for the financial support. 
\end{acknowledgments}

\appendix

\section{Supplementary material}

\subsection{Minimization process}

From the thermodynamic point of view, the favorable wetting state is the one that has the smallest energy. We have shown  in section \ref{teorico_pillar} the energy equations associated to the two wetting states, CB and W, both for pillared and porous surfaces. In this section we exemplify the procedure of finding the state with smallest energy for pillared surface, but it can also be applied for the porous surface.

\begin{figure}[H]
\centering
\includegraphics[width=.8\linewidth]{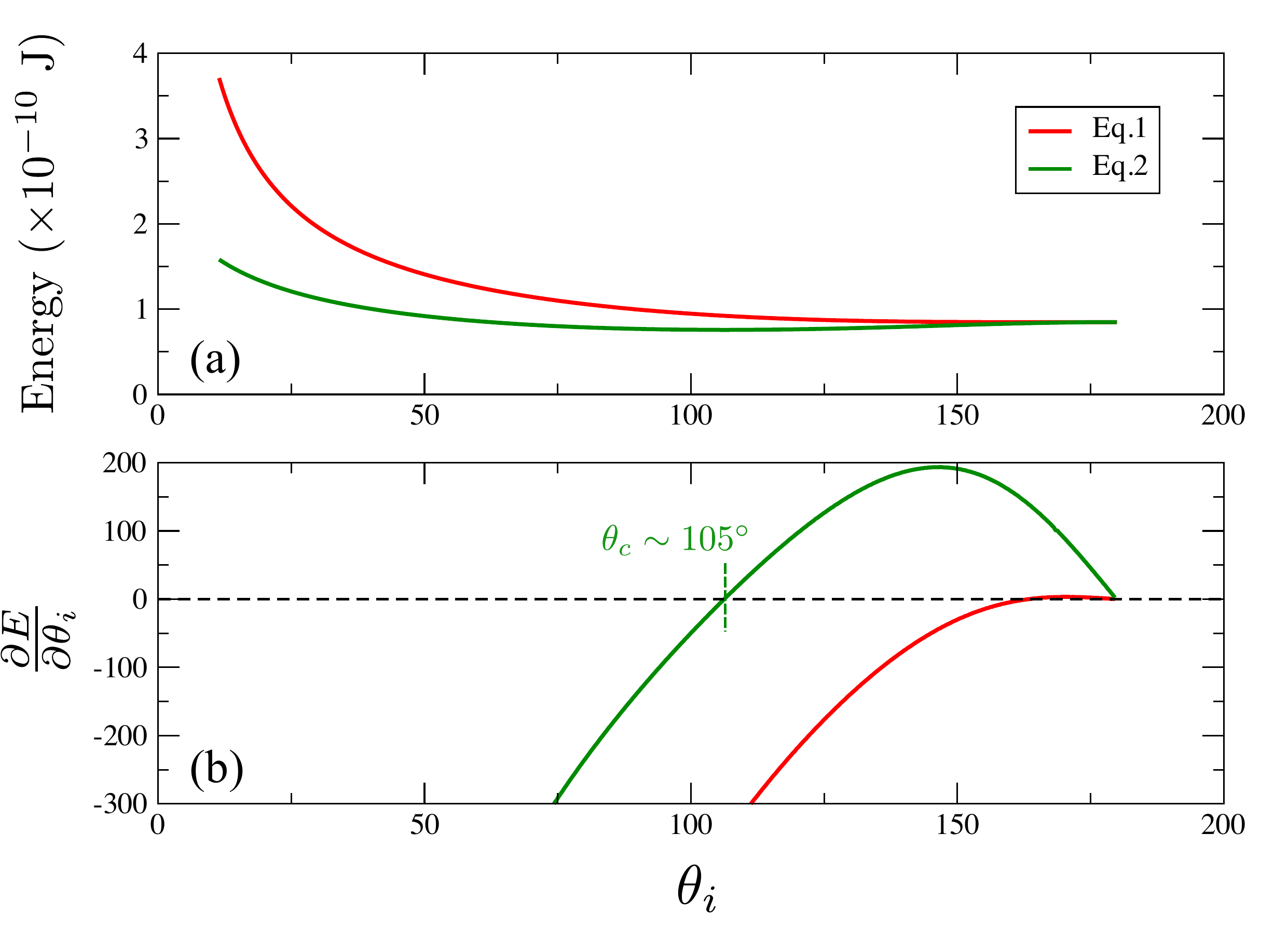}
\caption{{\bf (a)} red curve is Eq.(\ref{cb_pilares}) for the CB state and in green is Eq.(\ref{w_pilares}) for the W state. {\bf (b)} derivative of energy as a function of the angle parameter $\theta_i$. Dotted line shows where $\Delta E^{\rm CB}$ and $\Delta E^{\rm W}$ have their minimum value and the contact angle associated with it.}
\label{Minimo}
\end{figure}

For a pillared surface the process is the following: i) fix the surface parameters ($\aaa, \hh, \ww$) and the volume  $V_0=\frac{4}{3}\pi R_0^3$  of the droplet. ii) We solve a cubic equation to obtain the radius of the droplet in each state, CB or W. iii) We  vary the contact angle  $\theta_i$ and use Eqs.(\ref{cb_pilares}) and (\ref{w_pilares}) to calculate $\Delta E^{\rm CB}$ and $\Delta E^{\rm W}$. An example of these curves are shown in Fig.(\ref{Minimo}-a). iv) By deriving the energies curves in relation to $\theta_i$, we  find the global minimum energy for CB, $\Delta E^{\rm CB}_{\rm min}$ and for W state, $\Delta E^{\rm W}_{\rm min}$, as shown in Fig.(\ref{Minimo}-b).  We then compare both values to define which one is the thermodynamic stable state. For this set of surface parameters ($\aaa=14 \, \mu$m and $\hh=10 \, \mu$m) shown in Fig.(\ref{Minimo}), ~$\Delta E^{\rm W}_{\rm min} < \Delta E^{\rm CB}_{\rm min}$, which indicates that  W is the stable state. The contact angle of the droplet is the one calculated in the point of the lowest energy, which in this case is $\ttc = 105^{\circ}$. 

\subsection{Choosing numerical parameters}

In this section we analyze with further details the 4-state Potts model. To evolve the system we use the Metropolis algorithm, which consist on selecting a random site at the interface (between gas-water, gas-oil or water-oil) and attempt a change of the state spin. This trial is accepted if $\Delta H < 0$ and, if $\Delta H \geq 0 $, the flip is accepted with a probability of ${\rm e}^{- \beta \Delta H}$, where $\beta = 1/T$ and $T$ is an effective temperature of the CPM and acts a noise parameters that allows the phase space to be explored. 

The choice of the parameter $T$ is an important step in the simulation. We look for a values of $T$ that fulfill two conditions: a) $T$ cannot be large enough to evaporate the liquid (in this case the volume would not be maintained) and 
b) it cannot be too small, because it would in practice freeze the dynamics. In other words, the acceptance rate would become so small that it would be necessary a long-time simulations in order to observe any significant change.

\begin{figure}[H]
\centering
\includegraphics[width=.6\linewidth]{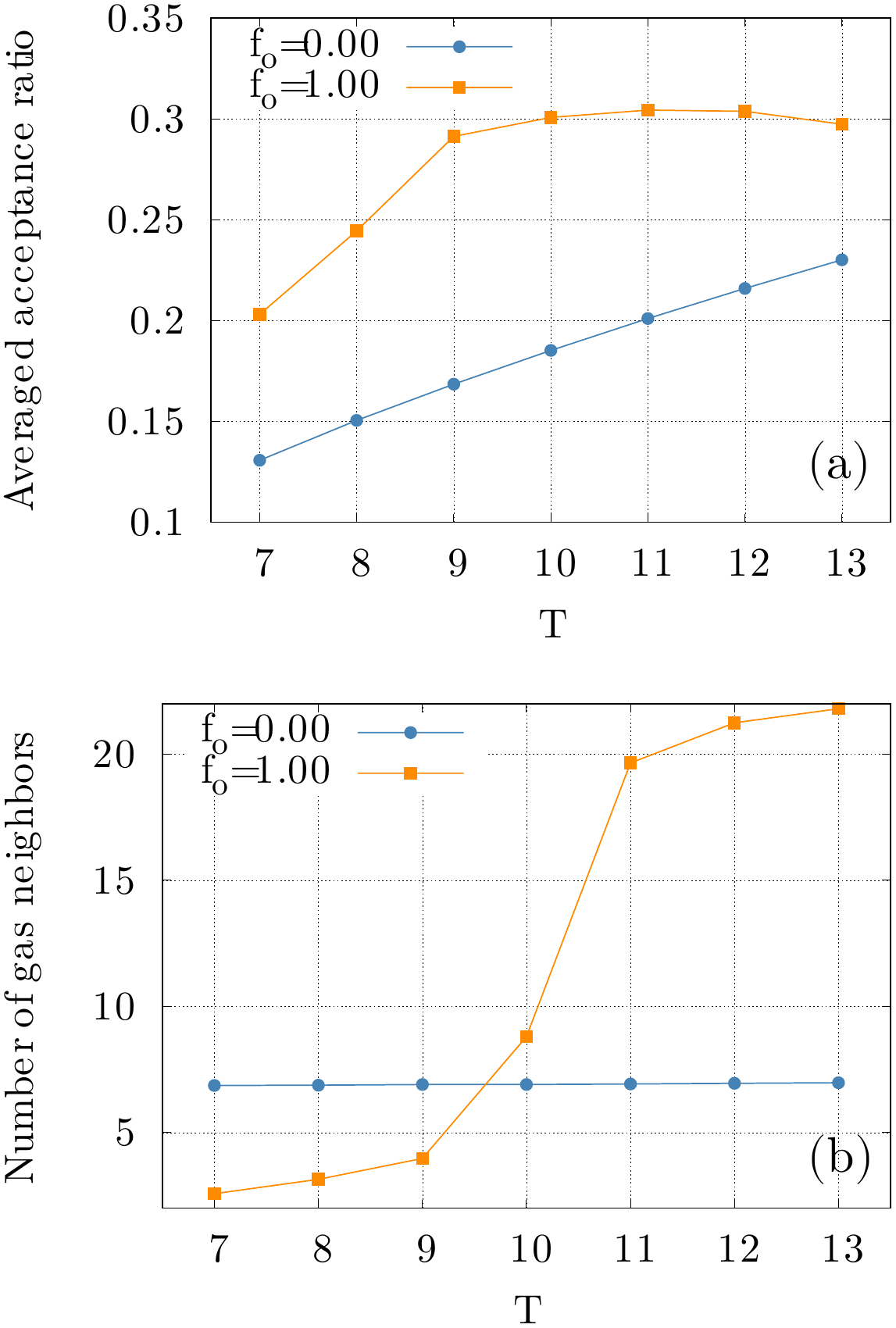}
\caption{{\bf (a)} average acceptance ratio as a function of parameter $T$ for a droplet of water ($f_o=0$) and a droplet of oil ($f_o=1$. {\bf (b)} average number of neighbors of a liquid spin that have a gas neighbour.}
\label{T_AccViz}
\end{figure}

Fig.(\ref{T_AccViz}-a) shows the average acceptance ratio for a droplet with pure water ($f_o=0$) and a droplet with pure oil ($f_o=1$) as a function of T. In Fig.(\ref{T_AccViz}-b) we show the average number of gas neighbors of the liquid site (water or oil). With these figures we reach the following conclusion: i) for $T>10$ the acceptance rate reaches a plateau of $\sim 30 \% $ for a droplet of oil. ii) in this same range the number of gas sites neighbors of a liquid site increases rapidly, meaning that each liquid site has only gas neighbors and indicating that the droplet had evaporated. iii) for $T < 7$ the acceptance ratio of a droplet of water is less than 10\%. An acceptance ratio this low indicates that a long-time simulation is necessary to observe change in the droplet. Therefore, with these conclusions we define $T=9$ as the parameter that fulfill the conditions "a" and "b" explained above.

\begin{figure}[H]
\centering
\includegraphics[width=.8\linewidth]{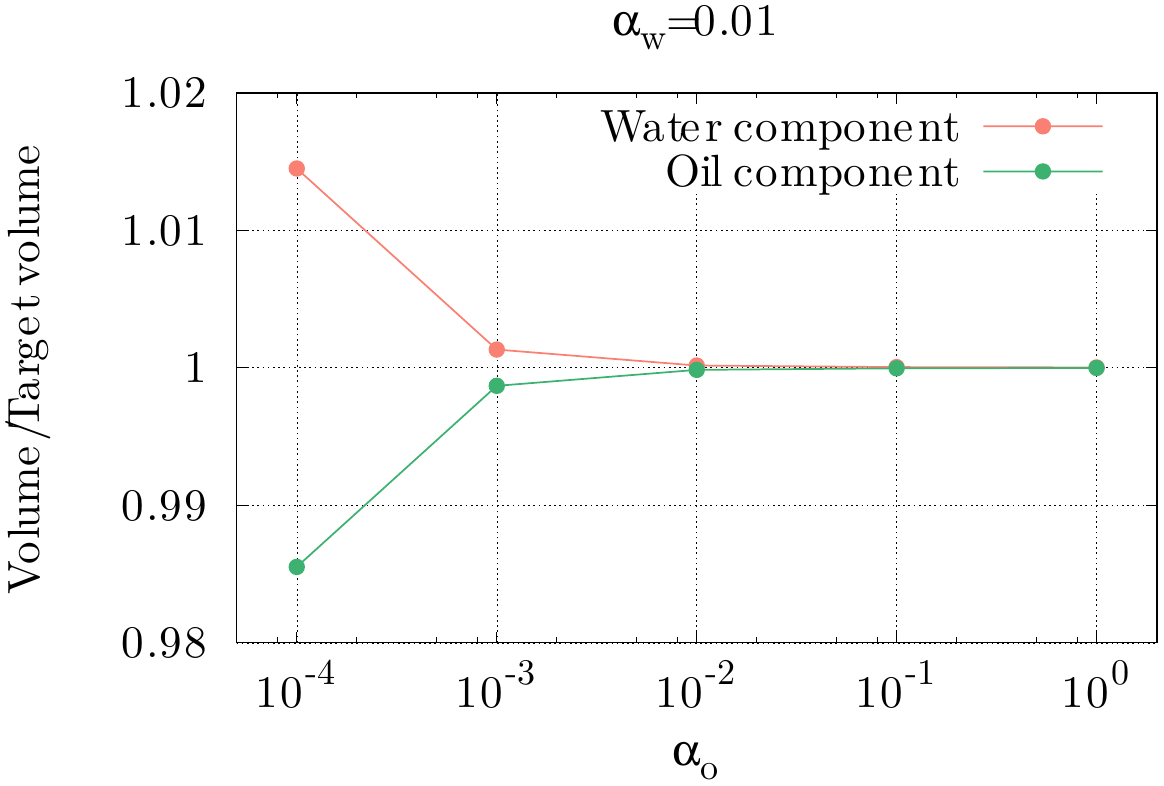}
\caption{Green (coral) curve is the volume of the water (oil) component of the droplet divided by its target value for a drop with $f_0=0.50$.}
\label{alfaW}
\end{figure}

The next step in the simulation is adjusting the parameters $\alpha_w$ and $\alpha_o$ from Eq.(\ref{hamil}). These are constraint parameters that forces the droplet to have a specific volume. If $f_o=0.50$, we are initializing a droplet with equal parts of water and oil. To decide the values of the parameters we fix a $\alpha_w$ and then range $\alpha_o$. In Fig.(\ref{alfaW}) we show the volume of each component divided by its target value as a function of $\alpha_o$ with $\alpha_w=0.01$. As we can see from this figure, lower values $\alpha_o$ allow for some fluctuations on the volume of both components and, since we are studying a droplet with constant volume, we will use $\alpha_w=0.01$ and $\alpha_o=0.01$.

\subsection{Results for the surfaces with $\hh=5 \mu$m}

In this section we show the simulations results for the pillared surface and porous surface with $\hh=5 \mu$m.

\begin{figure}[H]
\centering
\includegraphics[width=1\linewidth]{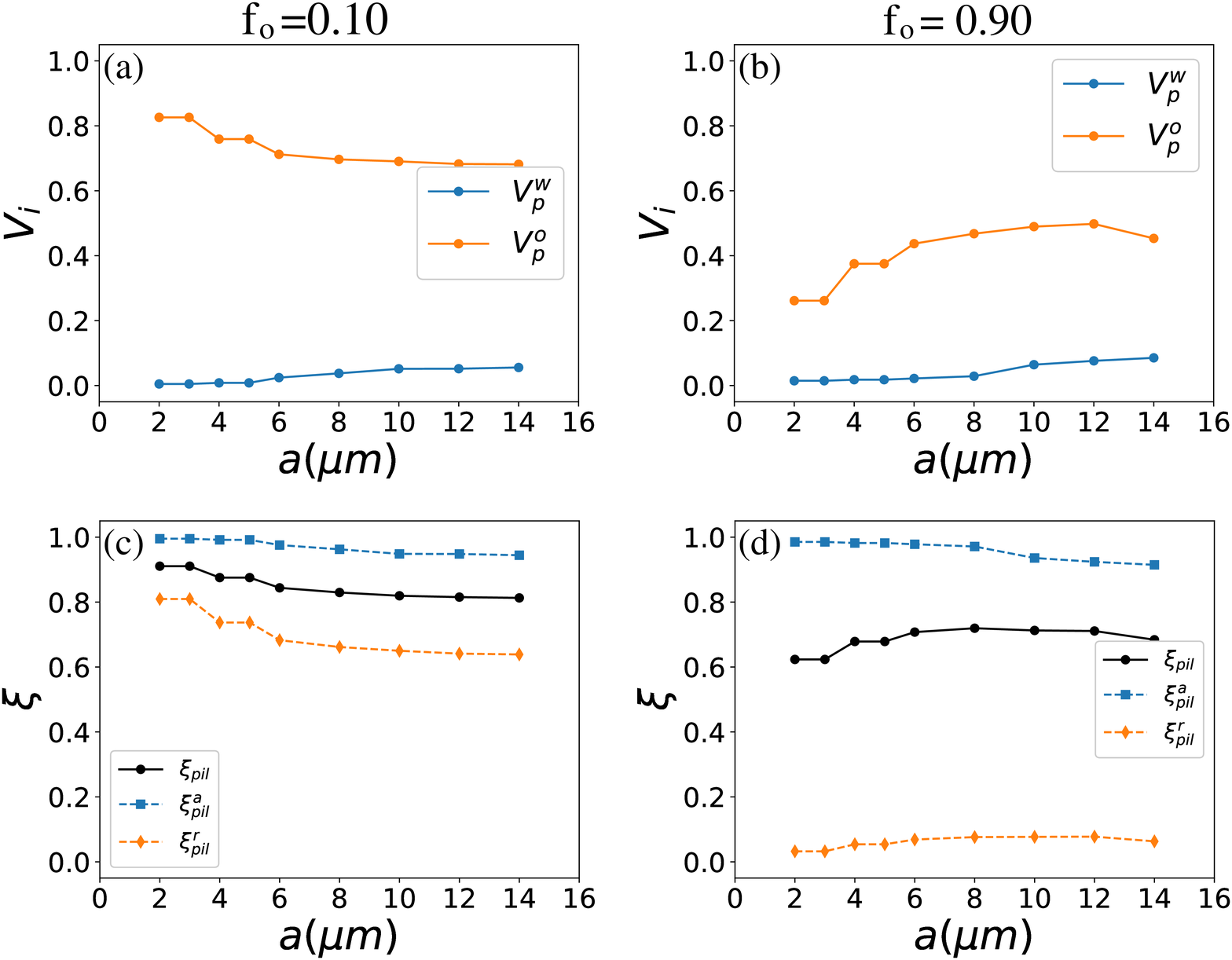}
\caption{{\bf Results for pillared substrates.} Interpillar volume of water, V$^w_p$, and oil, V$^o_p$, as a function of the pillar distance $\aaa$ for (a) $f_o = 0.10$ and (b) $f_o = 0.90$. Blue color represent water and orange represent oil. $\xi^{\bf a}_{\bf por}$, $\xi^{\bf r}_{\bf por}$ and  $\xi_{\bf por}$ as a function of interpillar distance $\aaa$ for (c)$f_o=0.10$ and (d)$f_o=0.90$ }
\label{comp_pillar}
\end{figure}

\begin{figure}[H]
\centering
\includegraphics[width=1\linewidth]{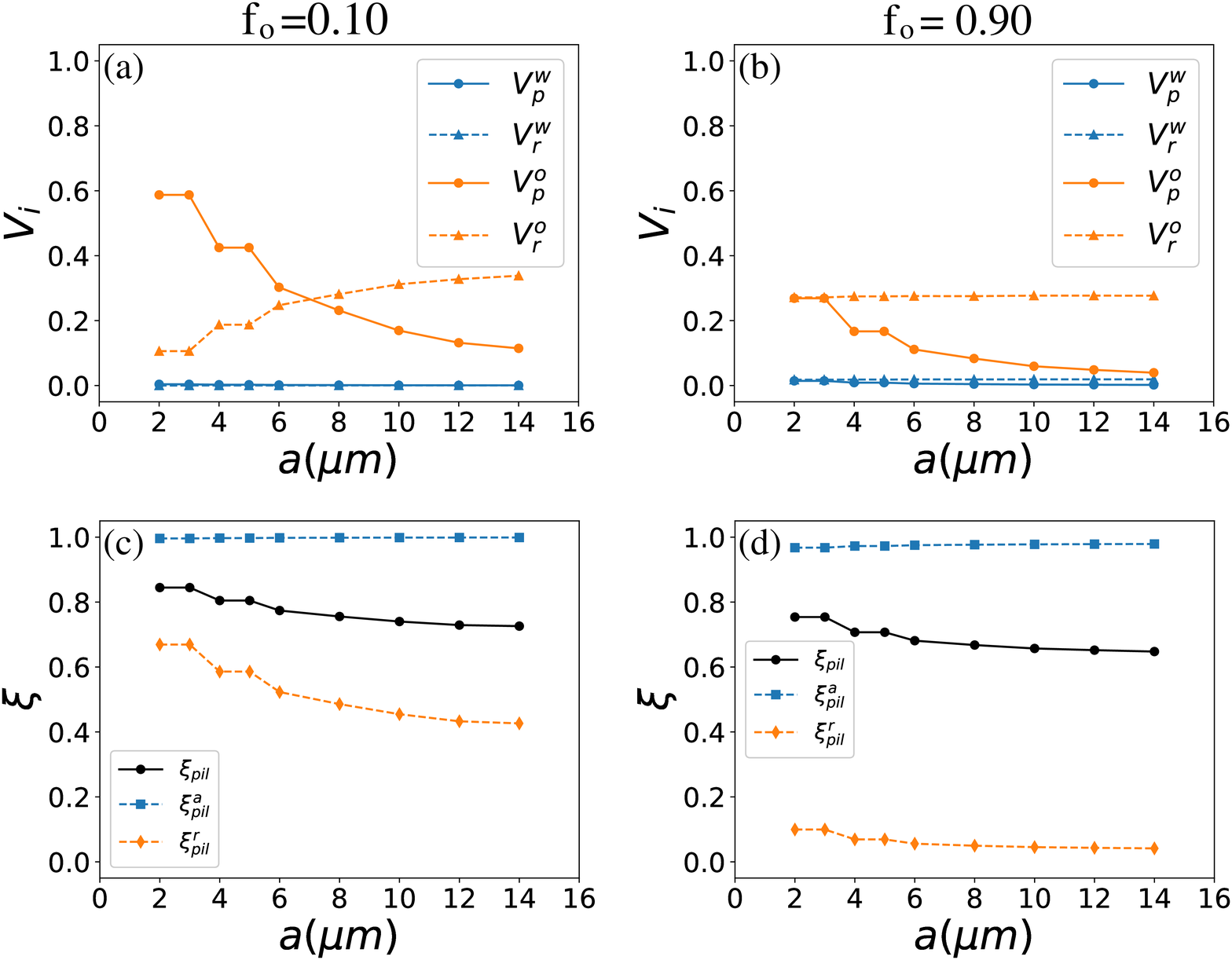}
\caption{{\bf Results for porous substrates.} Interporous volume  and reservoir volume of water, V$^w_p$ and V$^w_r$ , and oil, V$^o_p$ and V$^o_r$ , as a function of the porous distance $\aaa$ for (a) $f_o = 0.10$ and (b) $f_o = 0.90$. Blue color represent water and orange represent oil. $\xi^{\bf a}_{\bf por}$, $\xi^{\bf r}_{\bf por}$ and  $\xi_{\bf por}$ as a function of interporous distance $\aaa$ for (c)$f_o=0.10$ and (d)$f_o=0.90$}
\label{comp_poros}
\end{figure}

\newpage

\bibliography{bib_aip}

\begin{thebibliography}{47}%
\makeatletter
\providecommand \@ifxundefined [1]{%
 \@ifx{#1\undefined}
}%
\providecommand \@ifnum [1]{%
 \ifnum #1\expandafter \@firstoftwo
 \else \expandafter \@secondoftwo
 \fi
}%
\providecommand \@ifx [1]{%
 \ifx #1\expandafter \@firstoftwo
 \else \expandafter \@secondoftwo
 \fi
}%
\providecommand \natexlab [1]{#1}%
\providecommand \enquote  [1]{``#1''}%
\providecommand \bibnamefont  [1]{#1}%
\providecommand \bibfnamefont [1]{#1}%
\providecommand \citenamefont [1]{#1}%
\providecommand \href@noop [0]{\@secondoftwo}%
\providecommand \href [0]{\begingroup \@sanitize@url \@href}%
\providecommand \@href[1]{\@@startlink{#1}\@@href}%
\providecommand \@@href[1]{\endgroup#1\@@endlink}%
\providecommand \@sanitize@url [0]{\catcode `\\12\catcode `\$12\catcode
  `\&12\catcode `\#12\catcode `\^12\catcode `\_12\catcode `\%12\relax}%
\providecommand \@@startlink[1]{}%
\providecommand \@@endlink[0]{}%
\providecommand \url  [0]{\begingroup\@sanitize@url \@url }%
\providecommand \@url [1]{\endgroup\@href {#1}{\urlprefix }}%
\providecommand \urlprefix  [0]{URL }%
\providecommand \Eprint [0]{\href }%
\providecommand \doibase [0]{http://dx.doi.org/}%
\providecommand \selectlanguage [0]{\@gobble}%
\providecommand \bibinfo  [0]{\@secondoftwo}%
\providecommand \bibfield  [0]{\@secondoftwo}%
\providecommand \translation [1]{[#1]}%
\providecommand \BibitemOpen [0]{}%
\providecommand \bibitemStop [0]{}%
\providecommand \bibitemNoStop [0]{.\EOS\space}%
\providecommand \EOS [0]{\spacefactor3000\relax}%
\providecommand \BibitemShut  [1]{\csname bibitem#1\endcsname}%
\let\auto@bib@innerbib\@empty
\bibitem [{\citenamefont {Mekonnen}\ and\ \citenamefont
  {Hoekstra}(2016)}]{mek16}%
  \BibitemOpen
  \bibfield  {author} {\bibinfo {author} {\bibfnamefont {M.~M.}\ \bibnamefont
  {Mekonnen}}\ and\ \bibinfo {author} {\bibfnamefont {A.~Y.}\ \bibnamefont
  {Hoekstra}},\ }\bibfield  {title} {\enquote {\bibinfo {title} {Four billion
  people facing severe water scarcity},}\ }\href@noop {} {\bibfield  {journal}
  {\bibinfo  {journal} {Science advances}\ }\textbf {\bibinfo {volume} {2}},\
  \bibinfo {pages} {e1500323} (\bibinfo {year} {2016})}\BibitemShut {NoStop}%
\bibitem [{\citenamefont {Huang}\ \emph {et~al.}(2019)\citenamefont {Huang},
  \citenamefont {Joshi}, \citenamefont {Silva}, \citenamefont {Badam},\ and\
  \citenamefont {Yoshimura}}]{huang2019fabrication}%
  \BibitemOpen
  \bibfield  {author} {\bibinfo {author} {\bibfnamefont {H.}~\bibnamefont
  {Huang}}, \bibinfo {author} {\bibfnamefont {R.}~\bibnamefont {Joshi}},
  \bibinfo {author} {\bibfnamefont {K.~D.}\ \bibnamefont {Silva}}, \bibinfo
  {author} {\bibfnamefont {R.}~\bibnamefont {Badam}}, \ and\ \bibinfo {author}
  {\bibfnamefont {M.}~\bibnamefont {Yoshimura}},\ }\bibfield  {title} {\enquote
  {\bibinfo {title} {Fabrication of reduced graphene oxide membranes for water
  desalination},}\ }\href@noop {} {\bibfield  {journal} {\bibinfo  {journal}
  {Journal of membrane science}\ }\textbf {\bibinfo {volume} {572}},\ \bibinfo
  {pages} {12--19} (\bibinfo {year} {2019})}\BibitemShut {NoStop}%
\bibitem [{\citenamefont {Yang}\ \emph {et~al.}(2019)\citenamefont {Yang},
  \citenamefont {Yang}, \citenamefont {Liang}, \citenamefont {Gao},
  \citenamefont {Cheng}, \citenamefont {Li}, \citenamefont {Zou}, \citenamefont
  {Ma}, \citenamefont {Yuan},\ and\ \citenamefont {Duan}}]{yang2019large}%
  \BibitemOpen
  \bibfield  {author} {\bibinfo {author} {\bibfnamefont {Y.}~\bibnamefont
  {Yang}}, \bibinfo {author} {\bibfnamefont {X.}~\bibnamefont {Yang}}, \bibinfo
  {author} {\bibfnamefont {L.}~\bibnamefont {Liang}}, \bibinfo {author}
  {\bibfnamefont {Y.}~\bibnamefont {Gao}}, \bibinfo {author} {\bibfnamefont
  {H.}~\bibnamefont {Cheng}}, \bibinfo {author} {\bibfnamefont
  {X.}~\bibnamefont {Li}}, \bibinfo {author} {\bibfnamefont {M.}~\bibnamefont
  {Zou}}, \bibinfo {author} {\bibfnamefont {R.}~\bibnamefont {Ma}}, \bibinfo
  {author} {\bibfnamefont {Q.}~\bibnamefont {Yuan}}, \ and\ \bibinfo {author}
  {\bibfnamefont {X.}~\bibnamefont {Duan}},\ }\bibfield  {title} {\enquote
  {\bibinfo {title} {Large-area graphene-nanomesh/carbon-nanotube hybrid
  membranes for ionic and molecular nanofiltration},}\ }\href@noop {}
  {\bibfield  {journal} {\bibinfo  {journal} {Science}\ }\textbf {\bibinfo
  {volume} {364}},\ \bibinfo {pages} {1057--1062} (\bibinfo {year}
  {2019})}\BibitemShut {NoStop}%
\bibitem [{\citenamefont {Qian}\ \emph {et~al.}(2019)\citenamefont {Qian},
  \citenamefont {Zhang}, \citenamefont {Liu}, \citenamefont {Zhou},\ and\
  \citenamefont {Huang}}]{qian2019tuning}%
  \BibitemOpen
  \bibfield  {author} {\bibinfo {author} {\bibfnamefont {Y.}~\bibnamefont
  {Qian}}, \bibinfo {author} {\bibfnamefont {X.}~\bibnamefont {Zhang}},
  \bibinfo {author} {\bibfnamefont {C.}~\bibnamefont {Liu}}, \bibinfo {author}
  {\bibfnamefont {C.}~\bibnamefont {Zhou}}, \ and\ \bibinfo {author}
  {\bibfnamefont {A.}~\bibnamefont {Huang}},\ }\bibfield  {title} {\enquote
  {\bibinfo {title} {Tuning interlayer spacing of graphene oxide membranes with
  enhanced desalination performance},}\ }\href@noop {} {\bibfield  {journal}
  {\bibinfo  {journal} {Desalination}\ }\textbf {\bibinfo {volume} {460}},\
  \bibinfo {pages} {56--63} (\bibinfo {year} {2019})}\BibitemShut {NoStop}%
\bibitem [{\citenamefont {Xue}\ \emph {et~al.}(2014)\citenamefont {Xue},
  \citenamefont {Cao}, \citenamefont {Liu}, \citenamefont {Feng},\ and\
  \citenamefont {Jiang}}]{xue14}%
  \BibitemOpen
  \bibfield  {author} {\bibinfo {author} {\bibfnamefont {Z.}~\bibnamefont
  {Xue}}, \bibinfo {author} {\bibfnamefont {Y.}~\bibnamefont {Cao}}, \bibinfo
  {author} {\bibfnamefont {N.}~\bibnamefont {Liu}}, \bibinfo {author}
  {\bibfnamefont {L.}~\bibnamefont {Feng}}, \ and\ \bibinfo {author}
  {\bibfnamefont {L.}~\bibnamefont {Jiang}},\ }\bibfield  {title} {\enquote
  {\bibinfo {title} {Special wettable materials for oil/water separation},}\
  }\href@noop {} {\bibfield  {journal} {\bibinfo  {journal} {Journal of
  Materials Chemistry A}\ }\textbf {\bibinfo {volume} {2}},\ \bibinfo {pages}
  {2445--2460} (\bibinfo {year} {2014})}\BibitemShut {NoStop}%
\bibitem [{\citenamefont {Chan}\ \emph {et~al.}(2009)\citenamefont {Chan},
  \citenamefont {Chong}, \citenamefont {Law},\ and\ \citenamefont
  {Hassell}}]{chan09}%
  \BibitemOpen
  \bibfield  {author} {\bibinfo {author} {\bibfnamefont {Y.~J.}\ \bibnamefont
  {Chan}}, \bibinfo {author} {\bibfnamefont {M.~F.}\ \bibnamefont {Chong}},
  \bibinfo {author} {\bibfnamefont {C.~L.}\ \bibnamefont {Law}}, \ and\
  \bibinfo {author} {\bibfnamefont {D.~G.}\ \bibnamefont {Hassell}},\
  }\bibfield  {title} {\enquote {\bibinfo {title} {A review on
  anaerobic--aerobic treatment of industrial and municipal wastewater},}\
  }\href@noop {} {\bibfield  {journal} {\bibinfo  {journal} {Chemical
  Engineering Journal}\ }\textbf {\bibinfo {volume} {155}},\ \bibinfo {pages}
  {1--18} (\bibinfo {year} {2009})}\BibitemShut {NoStop}%
\bibitem [{\citenamefont {Padaki}\ \emph {et~al.}(2015)\citenamefont {Padaki},
  \citenamefont {Murali}, \citenamefont {Abdullah}, \citenamefont {Misdan},
  \citenamefont {Moslehyani}, \citenamefont {Kassim}, \citenamefont {Hilal},\
  and\ \citenamefont {Ismail}}]{pa15}%
  \BibitemOpen
  \bibfield  {author} {\bibinfo {author} {\bibfnamefont {M.}~\bibnamefont
  {Padaki}}, \bibinfo {author} {\bibfnamefont {R.~S.}\ \bibnamefont {Murali}},
  \bibinfo {author} {\bibfnamefont {M.~S.}\ \bibnamefont {Abdullah}}, \bibinfo
  {author} {\bibfnamefont {N.}~\bibnamefont {Misdan}}, \bibinfo {author}
  {\bibfnamefont {A.}~\bibnamefont {Moslehyani}}, \bibinfo {author}
  {\bibfnamefont {M.}~\bibnamefont {Kassim}}, \bibinfo {author} {\bibfnamefont
  {N.}~\bibnamefont {Hilal}}, \ and\ \bibinfo {author} {\bibfnamefont
  {A.}~\bibnamefont {Ismail}},\ }\bibfield  {title} {\enquote {\bibinfo {title}
  {Membrane technology enhancement in oil--water separation. a review},}\
  }\href@noop {} {\bibfield  {journal} {\bibinfo  {journal} {Desalination}\
  }\textbf {\bibinfo {volume} {357}},\ \bibinfo {pages} {197--207} (\bibinfo
  {year} {2015})}\BibitemShut {NoStop}%
\bibitem [{\citenamefont {Chen}\ \emph {et~al.}(2019)\citenamefont {Chen},
  \citenamefont {Weng}, \citenamefont {Mahmood}, \citenamefont {Chen},\ and\
  \citenamefont {Wang}}]{chen19}%
  \BibitemOpen
  \bibfield  {author} {\bibinfo {author} {\bibfnamefont {C.}~\bibnamefont
  {Chen}}, \bibinfo {author} {\bibfnamefont {D.}~\bibnamefont {Weng}}, \bibinfo
  {author} {\bibfnamefont {A.}~\bibnamefont {Mahmood}}, \bibinfo {author}
  {\bibfnamefont {S.}~\bibnamefont {Chen}}, \ and\ \bibinfo {author}
  {\bibfnamefont {J.}~\bibnamefont {Wang}},\ }\bibfield  {title} {\enquote
  {\bibinfo {title} {Separation mechanism and construction of surfaces with
  special wettability for oil/water separation},}\ }\href@noop {} {\bibfield
  {journal} {\bibinfo  {journal} {ACS applied materials \& interfaces}\
  }\textbf {\bibinfo {volume} {11}},\ \bibinfo {pages} {11006--11027} (\bibinfo
  {year} {2019})}\BibitemShut {NoStop}%
\bibitem [{\citenamefont {Kota}\ \emph {et~al.}(2012)\citenamefont {Kota},
  \citenamefont {Kwon}, \citenamefont {Choi}, \citenamefont {Mabry},\ and\
  \citenamefont {Tuteja}}]{kota12}%
  \BibitemOpen
  \bibfield  {author} {\bibinfo {author} {\bibfnamefont {A.~K.}\ \bibnamefont
  {Kota}}, \bibinfo {author} {\bibfnamefont {G.}~\bibnamefont {Kwon}}, \bibinfo
  {author} {\bibfnamefont {W.}~\bibnamefont {Choi}}, \bibinfo {author}
  {\bibfnamefont {J.~M.}\ \bibnamefont {Mabry}}, \ and\ \bibinfo {author}
  {\bibfnamefont {A.}~\bibnamefont {Tuteja}},\ }\bibfield  {title} {\enquote
  {\bibinfo {title} {Hygro-responsive membranes for effective oil--water
  separation},}\ }\href@noop {} {\bibfield  {journal} {\bibinfo  {journal}
  {Nature communications}\ }\textbf {\bibinfo {volume} {3}},\ \bibinfo {pages}
  {1--8} (\bibinfo {year} {2012})}\BibitemShut {NoStop}%
\bibitem [{\citenamefont {Cheryan}\ and\ \citenamefont
  {Rajagopalan}(1998)}]{cher98}%
  \BibitemOpen
  \bibfield  {author} {\bibinfo {author} {\bibfnamefont {M.}~\bibnamefont
  {Cheryan}}\ and\ \bibinfo {author} {\bibfnamefont {N.}~\bibnamefont
  {Rajagopalan}},\ }\bibfield  {title} {\enquote {\bibinfo {title} {Membrane
  processing of oily streams. wastewater treatment and waste reduction},}\
  }\href@noop {} {\bibfield  {journal} {\bibinfo  {journal} {Journal of
  membrane science}\ }\textbf {\bibinfo {volume} {151}},\ \bibinfo {pages}
  {13--28} (\bibinfo {year} {1998})}\BibitemShut {NoStop}%
\bibitem [{\citenamefont {Zeevalkink}\ and\ \citenamefont
  {Brunsmann}(1983)}]{zee83}%
  \BibitemOpen
  \bibfield  {author} {\bibinfo {author} {\bibfnamefont {J.}~\bibnamefont
  {Zeevalkink}}\ and\ \bibinfo {author} {\bibfnamefont {J.}~\bibnamefont
  {Brunsmann}},\ }\bibfield  {title} {\enquote {\bibinfo {title} {Oil removal
  from water in parallel plate gravity-type separators},}\ }\href@noop {}
  {\bibfield  {journal} {\bibinfo  {journal} {Water research}\ }\textbf
  {\bibinfo {volume} {17}},\ \bibinfo {pages} {365--373} (\bibinfo {year}
  {1983})}\BibitemShut {NoStop}%
\bibitem [{\citenamefont {Sun}\ \emph {et~al.}(1998)\citenamefont {Sun},
  \citenamefont {Duan}, \citenamefont {Li},\ and\ \citenamefont
  {Zhou}}]{sun98}%
  \BibitemOpen
  \bibfield  {author} {\bibinfo {author} {\bibfnamefont {D.}~\bibnamefont
  {Sun}}, \bibinfo {author} {\bibfnamefont {X.}~\bibnamefont {Duan}}, \bibinfo
  {author} {\bibfnamefont {W.}~\bibnamefont {Li}}, \ and\ \bibinfo {author}
  {\bibfnamefont {D.}~\bibnamefont {Zhou}},\ }\bibfield  {title} {\enquote
  {\bibinfo {title} {Demulsification of water-in-oil emulsion by using porous
  glass membrane},}\ }\href@noop {} {\bibfield  {journal} {\bibinfo  {journal}
  {Journal of membrane science}\ }\textbf {\bibinfo {volume} {146}},\ \bibinfo
  {pages} {65--72} (\bibinfo {year} {1998})}\BibitemShut {NoStop}%
\bibitem [{\citenamefont {Cambiella}\ \emph {et~al.}(2006)\citenamefont
  {Cambiella}, \citenamefont {Benito}, \citenamefont {Pazos},\ and\
  \citenamefont {Coca}}]{cam06}%
  \BibitemOpen
  \bibfield  {author} {\bibinfo {author} {\bibfnamefont {A.}~\bibnamefont
  {Cambiella}}, \bibinfo {author} {\bibfnamefont {J.}~\bibnamefont {Benito}},
  \bibinfo {author} {\bibfnamefont {C.}~\bibnamefont {Pazos}}, \ and\ \bibinfo
  {author} {\bibfnamefont {J.}~\bibnamefont {Coca}},\ }\bibfield  {title}
  {\enquote {\bibinfo {title} {Centrifugal separation efficiency in the
  treatment of waste emulsified oils},}\ }\href@noop {} {\bibfield  {journal}
  {\bibinfo  {journal} {Chemical Engineering Research and Design}\ }\textbf
  {\bibinfo {volume} {84}},\ \bibinfo {pages} {69--76} (\bibinfo {year}
  {2006})}\BibitemShut {NoStop}%
\bibitem [{\citenamefont {Comba}\ and\ \citenamefont {Kaiser}(1990)}]{com90}%
  \BibitemOpen
  \bibfield  {author} {\bibinfo {author} {\bibfnamefont {M.}~\bibnamefont
  {Comba}}\ and\ \bibinfo {author} {\bibfnamefont {K.}~\bibnamefont {Kaiser}},\
  }\bibfield  {title} {\enquote {\bibinfo {title} {Suspended particulate
  concentrations in the st. lawrence river (1985--1987) determined by
  centrifugation and filtration},}\ }\href@noop {} {\bibfield  {journal}
  {\bibinfo  {journal} {Science of The Total Environment}\ }\textbf {\bibinfo
  {volume} {97}},\ \bibinfo {pages} {191--206} (\bibinfo {year}
  {1990})}\BibitemShut {NoStop}%
\bibitem [{\citenamefont {Str{\o}m-Kristiansen}\ \emph
  {et~al.}(1995)\citenamefont {Str{\o}m-Kristiansen}, \citenamefont {Lewis},
  \citenamefont {Daling},\ and\ \citenamefont {Nordvik}}]{strom95}%
  \BibitemOpen
  \bibfield  {author} {\bibinfo {author} {\bibfnamefont {T.}~\bibnamefont
  {Str{\o}m-Kristiansen}}, \bibinfo {author} {\bibfnamefont {A.}~\bibnamefont
  {Lewis}}, \bibinfo {author} {\bibfnamefont {P.~S.}\ \bibnamefont {Daling}}, \
  and\ \bibinfo {author} {\bibfnamefont {A.~B.}\ \bibnamefont {Nordvik}},\
  }\bibfield  {title} {\enquote {\bibinfo {title} {Heat and chemical treatment
  of mechanically recovered w/o emulsions},}\ }\href@noop {} {\bibfield
  {journal} {\bibinfo  {journal} {Spill Science \& Technology Bulletin}\
  }\textbf {\bibinfo {volume} {2}},\ \bibinfo {pages} {133--141} (\bibinfo
  {year} {1995})}\BibitemShut {NoStop}%
\bibitem [{\citenamefont {Qu\'{e}r\'{e}}(2008)}]{Quere2008}%
  \BibitemOpen
  \bibfield  {author} {\bibinfo {author} {\bibfnamefont {D.}~\bibnamefont
  {Qu\'{e}r\'{e}}},\ }\bibfield  {title} {\enquote {\bibinfo {title} {{Wetting
  and roughness}},}\ }\href@noop {} {\bibfield  {journal} {\bibinfo  {journal}
  {Annu. Rev. Mater. Res.}\ }\textbf {\bibinfo {volume} {38}},\ \bibinfo
  {pages} {71--99} (\bibinfo {year} {2008})}\BibitemShut {NoStop}%
\bibitem [{\citenamefont {Wenzel}(1936)}]{we36}%
  \BibitemOpen
  \bibfield  {author} {\bibinfo {author} {\bibfnamefont {R.}~\bibnamefont
  {Wenzel}},\ }\bibfield  {title} {\enquote {\bibinfo {title} {Resistance of
  solid surfaces to wetting by water},}\ }\href@noop {} {\bibfield  {journal}
  {\bibinfo  {journal} {Industrial \& Engineering Chemistry}\ }\textbf
  {\bibinfo {volume} {28}},\ \bibinfo {pages} {988--994} (\bibinfo {year}
  {1936})}\BibitemShut {NoStop}%
\bibitem [{\citenamefont {Wenzel}(1949)}]{we49}%
  \BibitemOpen
  \bibfield  {author} {\bibinfo {author} {\bibfnamefont {R.}~\bibnamefont
  {Wenzel}},\ }\bibfield  {title} {\enquote {\bibinfo {title} {Surface
  roughness and contact angle.}}\ }\href@noop {} {\bibfield  {journal}
  {\bibinfo  {journal} {The Journal of Physical Chemistry}\ }\textbf {\bibinfo
  {volume} {53}},\ \bibinfo {pages} {1466--1467} (\bibinfo {year}
  {1949})}\BibitemShut {NoStop}%
\bibitem [{\citenamefont {Feng}\ \emph {et~al.}(2004)\citenamefont {Feng},
  \citenamefont {Zhang}, \citenamefont {Mai}, \citenamefont {Ma}, \citenamefont
  {Liu}, \citenamefont {Jiang},\ and\ \citenamefont {Zhu}}]{feng04}%
  \BibitemOpen
  \bibfield  {author} {\bibinfo {author} {\bibfnamefont {L.}~\bibnamefont
  {Feng}}, \bibinfo {author} {\bibfnamefont {Z.}~\bibnamefont {Zhang}},
  \bibinfo {author} {\bibfnamefont {Z.}~\bibnamefont {Mai}}, \bibinfo {author}
  {\bibfnamefont {Y.}~\bibnamefont {Ma}}, \bibinfo {author} {\bibfnamefont
  {B.}~\bibnamefont {Liu}}, \bibinfo {author} {\bibfnamefont {L.}~\bibnamefont
  {Jiang}}, \ and\ \bibinfo {author} {\bibfnamefont {D.}~\bibnamefont {Zhu}},\
  }\bibfield  {title} {\enquote {\bibinfo {title} {A super-hydrophobic and
  super-oleophilic coating mesh film for the separation of oil and water},}\
  }\href@noop {} {\bibfield  {journal} {\bibinfo  {journal} {Angewandte Chemie
  International Edition}\ }\textbf {\bibinfo {volume} {43}},\ \bibinfo {pages}
  {2012--2014} (\bibinfo {year} {2004})}\BibitemShut {NoStop}%
\bibitem [{\citenamefont {Gui}\ \emph {et~al.}(2010)\citenamefont {Gui},
  \citenamefont {Wei}, \citenamefont {Wang}, \citenamefont {Cao}, \citenamefont
  {Zhu}, \citenamefont {Jia}, \citenamefont {Shu},\ and\ \citenamefont
  {Wu}}]{gui10}%
  \BibitemOpen
  \bibfield  {author} {\bibinfo {author} {\bibfnamefont {X.}~\bibnamefont
  {Gui}}, \bibinfo {author} {\bibfnamefont {J.}~\bibnamefont {Wei}}, \bibinfo
  {author} {\bibfnamefont {K.}~\bibnamefont {Wang}}, \bibinfo {author}
  {\bibfnamefont {A.}~\bibnamefont {Cao}}, \bibinfo {author} {\bibfnamefont
  {H.}~\bibnamefont {Zhu}}, \bibinfo {author} {\bibfnamefont {Y.}~\bibnamefont
  {Jia}}, \bibinfo {author} {\bibfnamefont {Q.}~\bibnamefont {Shu}}, \ and\
  \bibinfo {author} {\bibfnamefont {D.}~\bibnamefont {Wu}},\ }\bibfield
  {title} {\enquote {\bibinfo {title} {Carbon nanotube sponges},}\ }\href@noop
  {} {\bibfield  {journal} {\bibinfo  {journal} {Advanced materials}\ }\textbf
  {\bibinfo {volume} {22}},\ \bibinfo {pages} {617--621} (\bibinfo {year}
  {2010})}\BibitemShut {NoStop}%
\bibitem [{\citenamefont {Cortese}\ \emph {et~al.}(2014)\citenamefont
  {Cortese}, \citenamefont {Caschera}, \citenamefont {Federici}, \citenamefont
  {Ingo},\ and\ \citenamefont {Gigli}}]{cortese14}%
  \BibitemOpen
  \bibfield  {author} {\bibinfo {author} {\bibfnamefont {B.}~\bibnamefont
  {Cortese}}, \bibinfo {author} {\bibfnamefont {D.}~\bibnamefont {Caschera}},
  \bibinfo {author} {\bibfnamefont {F.}~\bibnamefont {Federici}}, \bibinfo
  {author} {\bibfnamefont {G.~M.}\ \bibnamefont {Ingo}}, \ and\ \bibinfo
  {author} {\bibfnamefont {G.}~\bibnamefont {Gigli}},\ }\bibfield  {title}
  {\enquote {\bibinfo {title} {Superhydrophobic fabrics for oil--water
  separation through a diamond like carbon (dlc) coating},}\ }\href@noop {}
  {\bibfield  {journal} {\bibinfo  {journal} {Journal of Materials Chemistry
  A}\ }\textbf {\bibinfo {volume} {2}},\ \bibinfo {pages} {6781--6789}
  (\bibinfo {year} {2014})}\BibitemShut {NoStop}%
\bibitem [{\citenamefont {Zhan}\ \emph {et~al.}(2020)\citenamefont {Zhan},
  \citenamefont {He}, \citenamefont {j.~Hu}, \citenamefont {Zhao},
  \citenamefont {Zeng}, \citenamefont {Zhou}, \citenamefont {Zhang},\ and\
  \citenamefont {Sengupta}}]{zhan20}%
  \BibitemOpen
  \bibfield  {author} {\bibinfo {author} {\bibfnamefont {Y.}~\bibnamefont
  {Zhan}}, \bibinfo {author} {\bibfnamefont {S.}~\bibnamefont {He}}, \bibinfo
  {author} {\bibnamefont {j.~Hu}}, \bibinfo {author} {\bibfnamefont
  {S.}~\bibnamefont {Zhao}}, \bibinfo {author} {\bibfnamefont {G.}~\bibnamefont
  {Zeng}}, \bibinfo {author} {\bibfnamefont {.}~\bibnamefont {Zhou}}, \bibinfo
  {author} {\bibfnamefont {G.}~\bibnamefont {Zhang}}, \ and\ \bibinfo {author}
  {\bibfnamefont {A.}~\bibnamefont {Sengupta}},\ }\bibfield  {title} {\enquote
  {\bibinfo {title} {Robust super-hydrophobic/super-oleophilic sandwich-like
  uio-66-f4@ rgo composites for efficient and multitasking oil/water separation
  applications},}\ }\href@noop {} {\bibfield  {journal} {\bibinfo  {journal}
  {Journal of hazardous materials}\ }\textbf {\bibinfo {volume} {388}},\
  \bibinfo {pages} {121752} (\bibinfo {year} {2020})}\BibitemShut {NoStop}%
\bibitem [{\citenamefont {Yang}\ \emph {et~al.}(2012)\citenamefont {Yang},
  \citenamefont {Zhang}, \citenamefont {Xu}, \citenamefont {Zhu}, \citenamefont
  {Men},\ and\ \citenamefont {Zhou}}]{yang12}%
  \BibitemOpen
  \bibfield  {author} {\bibinfo {author} {\bibfnamefont {J.}~\bibnamefont
  {Yang}}, \bibinfo {author} {\bibfnamefont {Z.}~\bibnamefont {Zhang}},
  \bibinfo {author} {\bibfnamefont {X.}~\bibnamefont {Xu}}, \bibinfo {author}
  {\bibfnamefont {X.}~\bibnamefont {Zhu}}, \bibinfo {author} {\bibfnamefont
  {X.}~\bibnamefont {Men}}, \ and\ \bibinfo {author} {\bibfnamefont
  {X.}~\bibnamefont {Zhou}},\ }\bibfield  {title} {\enquote {\bibinfo {title}
  {Superhydrophilic--superoleophobic coatings},}\ }\href@noop {} {\bibfield
  {journal} {\bibinfo  {journal} {Journal of Materials Chemistry}\ }\textbf
  {\bibinfo {volume} {22}},\ \bibinfo {pages} {2834--2837} (\bibinfo {year}
  {2012})}\BibitemShut {NoStop}%
\bibitem [{\citenamefont {Tuteja}\ \emph {et~al.}(2007)\citenamefont {Tuteja},
  \citenamefont {Choi}, \citenamefont {Ma}, \citenamefont {Mabry},
  \citenamefont {Mazzella}, \citenamefont {Rutledge}, \citenamefont
  {McKinley},\ and\ \citenamefont {Cohen}}]{tuteja07}%
  \BibitemOpen
  \bibfield  {author} {\bibinfo {author} {\bibfnamefont {A.}~\bibnamefont
  {Tuteja}}, \bibinfo {author} {\bibfnamefont {W.}~\bibnamefont {Choi}},
  \bibinfo {author} {\bibfnamefont {M.}~\bibnamefont {Ma}}, \bibinfo {author}
  {\bibfnamefont {J.~M.}\ \bibnamefont {Mabry}}, \bibinfo {author}
  {\bibfnamefont {S.}~\bibnamefont {Mazzella}}, \bibinfo {author}
  {\bibfnamefont {G.}~\bibnamefont {Rutledge}}, \bibinfo {author}
  {\bibfnamefont {G.}~\bibnamefont {McKinley}}, \ and\ \bibinfo {author}
  {\bibfnamefont {R.~E.}\ \bibnamefont {Cohen}},\ }\bibfield  {title} {\enquote
  {\bibinfo {title} {Designing superoleophobic surfaces},}\ }\href@noop {}
  {\bibfield  {journal} {\bibinfo  {journal} {Science}\ }\textbf {\bibinfo
  {volume} {318}},\ \bibinfo {pages} {1618--1622} (\bibinfo {year}
  {2007})}\BibitemShut {NoStop}%
\bibitem [{\citenamefont {Ahuja}\ \emph {et~al.}(2008)\citenamefont {Ahuja},
  \citenamefont {Taylor}, \citenamefont {Lifton}, \citenamefont {Sidorenko},
  \citenamefont {Salamon}, \citenamefont {Lobaton}, \citenamefont {Kolodner},\
  and\ \citenamefont {Krupenkin}}]{ahuja08}%
  \BibitemOpen
  \bibfield  {author} {\bibinfo {author} {\bibfnamefont {A.}~\bibnamefont
  {Ahuja}}, \bibinfo {author} {\bibfnamefont {J.}~\bibnamefont {Taylor}},
  \bibinfo {author} {\bibfnamefont {V.}~\bibnamefont {Lifton}}, \bibinfo
  {author} {\bibfnamefont {A.}~\bibnamefont {Sidorenko}}, \bibinfo {author}
  {\bibfnamefont {T.}~\bibnamefont {Salamon}}, \bibinfo {author} {\bibfnamefont
  {E.}~\bibnamefont {Lobaton}}, \bibinfo {author} {\bibfnamefont
  {P.}~\bibnamefont {Kolodner}}, \ and\ \bibinfo {author} {\bibfnamefont
  {T.}~\bibnamefont {Krupenkin}},\ }\bibfield  {title} {\enquote {\bibinfo
  {title} {Nanonails: A simple geometrical approach to electrically tunable
  superlyophobic surfaces},}\ }\href@noop {} {\bibfield  {journal} {\bibinfo
  {journal} {Langmuir}\ }\textbf {\bibinfo {volume} {24}},\ \bibinfo {pages}
  {9--14} (\bibinfo {year} {2008})}\BibitemShut {NoStop}%
\bibitem [{\citenamefont {Sbragaglia}\ \emph {et~al.}(2007)\citenamefont
  {Sbragaglia}, \citenamefont {Peters}, \citenamefont {Pirat}, \citenamefont
  {Borkent}, \citenamefont {Lammertink}, \citenamefont {Wessling},\ and\
  \citenamefont {Lohse}}]{Sbragaglia2007}%
  \BibitemOpen
  \bibfield  {author} {\bibinfo {author} {\bibfnamefont {M.}~\bibnamefont
  {Sbragaglia}}, \bibinfo {author} {\bibfnamefont {A.}~\bibnamefont {Peters}},
  \bibinfo {author} {\bibfnamefont {C.}~\bibnamefont {Pirat}}, \bibinfo
  {author} {\bibfnamefont {B.}~\bibnamefont {Borkent}}, \bibinfo {author}
  {\bibfnamefont {R.}~\bibnamefont {Lammertink}}, \bibinfo {author}
  {\bibfnamefont {M.}~\bibnamefont {Wessling}}, \ and\ \bibinfo {author}
  {\bibfnamefont {D.}~\bibnamefont {Lohse}},\ }\bibfield  {title} {\enquote
  {\bibinfo {title} {{Spontaneous breakdown of superhydrophobicity}},}\
  }\href@noop {} {\bibfield  {journal} {\bibinfo  {journal} {Phys. Rev. Lett.}\
  }\textbf {\bibinfo {volume} {99}},\ \bibinfo {pages} {156001} (\bibinfo
  {year} {2007})}\BibitemShut {NoStop}%
\bibitem [{\citenamefont {Tsai}\ \emph {et~al.}(2010)\citenamefont {Tsai},
  \citenamefont {Lammertink}, \citenamefont {Wessling},\ and\ \citenamefont
  {Lohse}}]{Tsai2010}%
  \BibitemOpen
  \bibfield  {author} {\bibinfo {author} {\bibfnamefont {P.}~\bibnamefont
  {Tsai}}, \bibinfo {author} {\bibfnamefont {R.}~\bibnamefont {Lammertink}},
  \bibinfo {author} {\bibfnamefont {M.}~\bibnamefont {Wessling}}, \ and\
  \bibinfo {author} {\bibfnamefont {D.}~\bibnamefont {Lohse}},\ }\bibfield
  {title} {\enquote {\bibinfo {title} {Evaporation-triggered wetting transition
  for water droplets upon hydrophobic microstructures},}\ }\href@noop {}
  {\bibfield  {journal} {\bibinfo  {journal} {Physical review letters}\
  }\textbf {\bibinfo {volume} {104}},\ \bibinfo {pages} {116102} (\bibinfo
  {year} {2010})}\BibitemShut {NoStop}%
\bibitem [{\citenamefont {Shahraz}, \citenamefont {Borhan},\ and\ \citenamefont
  {Fichthorn}(2012)}]{Shahraz2012}%
  \BibitemOpen
  \bibfield  {author} {\bibinfo {author} {\bibfnamefont {A.}~\bibnamefont
  {Shahraz}}, \bibinfo {author} {\bibfnamefont {A.}~\bibnamefont {Borhan}}, \
  and\ \bibinfo {author} {\bibfnamefont {K.}~\bibnamefont {Fichthorn}},\
  }\bibfield  {title} {\enquote {\bibinfo {title} {{A theory for the
  morphological dependence of wetting on a physically patterned solid
  surface}},}\ }\href@noop {} {\bibfield  {journal} {\bibinfo  {journal}
  {Langmuir}\ }\textbf {\bibinfo {volume} {28}},\ \bibinfo {pages}
  {14227--14237} (\bibinfo {year} {2012})}\BibitemShut {NoStop}%
\bibitem [{\citenamefont {Fernandes}, \citenamefont {Vainstein},\ and\
  \citenamefont {Brito}(2015)}]{fernandes2015}%
  \BibitemOpen
  \bibfield  {author} {\bibinfo {author} {\bibfnamefont {H.}~\bibnamefont
  {Fernandes}}, \bibinfo {author} {\bibfnamefont {M.}~\bibnamefont
  {Vainstein}}, \ and\ \bibinfo {author} {\bibfnamefont {C.}~\bibnamefont
  {Brito}},\ }\bibfield  {title} {\enquote {\bibinfo {title} {Modeling of
  droplet evaporation on superhydrophobic surfaces},}\ }\href@noop {}
  {\bibfield  {journal} {\bibinfo  {journal} {Langmuir}\ }\textbf {\bibinfo
  {volume} {31}},\ \bibinfo {pages} {7652--7659} (\bibinfo {year}
  {2015})}\BibitemShut {NoStop}%
\bibitem [{\citenamefont {Silvestrini}\ and\ \citenamefont
  {Brito}(2017)}]{Silvestrini2017}%
  \BibitemOpen
  \bibfield  {author} {\bibinfo {author} {\bibfnamefont {M.}~\bibnamefont
  {Silvestrini}}\ and\ \bibinfo {author} {\bibfnamefont {C.}~\bibnamefont
  {Brito}},\ }\bibfield  {title} {\enquote {\bibinfo {title} {Wettability of
  reentrant surfaces: A global energy approach},}\ }\href@noop {} {\bibfield
  {journal} {\bibinfo  {journal} {Langmuir}\ }\textbf {\bibinfo {volume}
  {33}},\ \bibinfo {pages} {12535--12545} (\bibinfo {year} {2017})}\BibitemShut
  {NoStop}%
\bibitem [{\citenamefont {Lazzari}\ and\ \citenamefont {Brito}(2019)}]{laz19}%
  \BibitemOpen
  \bibfield  {author} {\bibinfo {author} {\bibfnamefont {D.}~\bibnamefont
  {Lazzari}}\ and\ \bibinfo {author} {\bibfnamefont {C.}~\bibnamefont
  {Brito}},\ }\bibfield  {title} {\enquote {\bibinfo {title} {Geometric and
  chemical nonuniformity may induce the stability of more than one wetting
  state in the same hydrophobic surface},}\ }\href@noop {} {\bibfield
  {journal} {\bibinfo  {journal} {Physical Review E}\ }\textbf {\bibinfo
  {volume} {99}},\ \bibinfo {pages} {032801} (\bibinfo {year}
  {2019})}\BibitemShut {NoStop}%
\bibitem [{\citenamefont {Martin}\ and\ \citenamefont
  {Bhushan}(2017)}]{martin17}%
  \BibitemOpen
  \bibfield  {author} {\bibinfo {author} {\bibfnamefont {S.}~\bibnamefont
  {Martin}}\ and\ \bibinfo {author} {\bibfnamefont {B.}~\bibnamefont
  {Bhushan}},\ }\bibfield  {title} {\enquote {\bibinfo {title} {Transparent,
  wear-resistant, superhydrophobic and superoleophobic poly
  (dimethylsiloxane)(pdms) surfaces},}\ }\href@noop {} {\bibfield  {journal}
  {\bibinfo  {journal} {Journal of colloid and interface science}\ }\textbf
  {\bibinfo {volume} {488}},\ \bibinfo {pages} {118--126} (\bibinfo {year}
  {2017})}\BibitemShut {NoStop}%
\bibitem [{\citenamefont {Wu}\ and\ \citenamefont {Hornof}(1999)}]{wu99}%
  \BibitemOpen
  \bibfield  {author} {\bibinfo {author} {\bibfnamefont {D.}~\bibnamefont
  {Wu}}\ and\ \bibinfo {author} {\bibfnamefont {V.}~\bibnamefont {Hornof}},\
  }\bibfield  {title} {\enquote {\bibinfo {title} {Dynamic interfacial tension
  in hexadecane/water systems containing ready-made and in-situ-formed
  surfactants},}\ }\href@noop {} {\bibfield  {journal} {\bibinfo  {journal}
  {Chemical Engineering Communications}\ }\textbf {\bibinfo {volume} {172}},\
  \bibinfo {pages} {85--106} (\bibinfo {year} {1999})}\BibitemShut {NoStop}%
\bibitem [{\citenamefont {Lopes}\ \emph {et~al.}(2013)\citenamefont {Lopes},
  \citenamefont {de~Oliveira}, \citenamefont {Ramos},\ and\ \citenamefont
  {Mombach}}]{Lopes2013}%
  \BibitemOpen
  \bibfield  {author} {\bibinfo {author} {\bibfnamefont {D.}~\bibnamefont
  {Lopes}}, \bibinfo {author} {\bibfnamefont {L.}~\bibnamefont {de~Oliveira}},
  \bibinfo {author} {\bibfnamefont {S.}~\bibnamefont {Ramos}}, \ and\ \bibinfo
  {author} {\bibfnamefont {J.}~\bibnamefont {Mombach}},\ }\bibfield  {title}
  {\enquote {\bibinfo {title} {{Cassie-Baxter to Wenzel state wetting
  transition: a 2D numerical simulation}},}\ }\href@noop {} {\bibfield
  {journal} {\bibinfo  {journal} {RSC Adv.}\ }\textbf {\bibinfo {volume} {3}},\
  \bibinfo {pages} {24530--24534} (\bibinfo {year} {2013})}\BibitemShut
  {NoStop}%
\bibitem [{\citenamefont {de~Oliveira}\ \emph {et~al.}(2011)\citenamefont
  {de~Oliveira}, \citenamefont {Lopes}, \citenamefont {Ramos},\ and\
  \citenamefont {Mombach}}]{Oliveira2011}%
  \BibitemOpen
  \bibfield  {author} {\bibinfo {author} {\bibfnamefont {L.}~\bibnamefont
  {de~Oliveira}}, \bibinfo {author} {\bibfnamefont {D.}~\bibnamefont {Lopes}},
  \bibinfo {author} {\bibfnamefont {S.}~\bibnamefont {Ramos}}, \ and\ \bibinfo
  {author} {\bibfnamefont {J.}~\bibnamefont {Mombach}},\ }\bibfield  {title}
  {\enquote {\bibinfo {title} {{Two-dimensional modeling of the
  superhydrophobic behavior of a liquid droplet sliding down a ramp of
  pillars}},}\ }\href@noop {} {\bibfield  {journal} {\bibinfo  {journal} {Soft
  Matter}\ }\textbf {\bibinfo {volume} {7}},\ \bibinfo {pages} {3763--3765}
  (\bibinfo {year} {2011})}\BibitemShut {NoStop}%
\bibitem [{\citenamefont {Mortazavi}, \citenamefont {D'Souza},\ and\
  \citenamefont {Nosonovsky}(2013)}]{Mortazavi2013}%
  \BibitemOpen
  \bibfield  {author} {\bibinfo {author} {\bibfnamefont {V.}~\bibnamefont
  {Mortazavi}}, \bibinfo {author} {\bibfnamefont {R.}~\bibnamefont {D'Souza}},
  \ and\ \bibinfo {author} {\bibfnamefont {M.}~\bibnamefont {Nosonovsky}},\
  }\bibfield  {title} {\enquote {\bibinfo {title} {{Study of contact angle
  hysteresis using the cellular Potts model}},}\ }\href@noop {} {\bibfield
  {journal} {\bibinfo  {journal} {Phys. Chem. Chem. Phys.}\ }\textbf {\bibinfo
  {volume} {15}},\ \bibinfo {pages} {2749--2756} (\bibinfo {year}
  {2013})}\BibitemShut {NoStop}%
\bibitem [{\citenamefont {Graner}\ and\ \citenamefont
  {Glazier}(1992)}]{Graner1992}%
  \BibitemOpen
  \bibfield  {author} {\bibinfo {author} {\bibfnamefont {F.}~\bibnamefont
  {Graner}}\ and\ \bibinfo {author} {\bibfnamefont {J.~A.}\ \bibnamefont
  {Glazier}},\ }\bibfield  {title} {\enquote {\bibinfo {title} {Simulation of
  biological cell sorting using a two-dimensional extended potts model},}\
  }\href@noop {} {\bibfield  {journal} {\bibinfo  {journal} {Phys. Rev. Lett.}\
  }\textbf {\bibinfo {volume} {69}},\ \bibinfo {pages} {2013--2017} (\bibinfo
  {year} {1992})}\BibitemShut {NoStop}%
\bibitem [{\citenamefont {Fortuna}\ \emph {et~al.}(2020)\citenamefont
  {Fortuna}, \citenamefont {Perrone}, \citenamefont {Krug}, \citenamefont
  {Susin}, \citenamefont {Belmonte}, \citenamefont {Thomas}, \citenamefont
  {Glazier},\ and\ \citenamefont {de~Almeida}}]{fortuna2020}%
  \BibitemOpen
  \bibfield  {author} {\bibinfo {author} {\bibfnamefont {I.}~\bibnamefont
  {Fortuna}}, \bibinfo {author} {\bibfnamefont {G.~C.}\ \bibnamefont
  {Perrone}}, \bibinfo {author} {\bibfnamefont {M.~S.}\ \bibnamefont {Krug}},
  \bibinfo {author} {\bibfnamefont {E.}~\bibnamefont {Susin}}, \bibinfo
  {author} {\bibfnamefont {J.~M.}\ \bibnamefont {Belmonte}}, \bibinfo {author}
  {\bibfnamefont {G.~L.}\ \bibnamefont {Thomas}}, \bibinfo {author}
  {\bibfnamefont {J.~A.}\ \bibnamefont {Glazier}}, \ and\ \bibinfo {author}
  {\bibfnamefont {R.~M.}\ \bibnamefont {de~Almeida}},\ }\bibfield  {title}
  {\enquote {\bibinfo {title} {Compucell3d simulations reproduce mesenchymal
  cell migration on flat substrates.}}\ }\href@noop {} {\bibfield  {journal}
  {\bibinfo  {journal} {Biophysical Journal}\ } (\bibinfo {year}
  {2020})}\BibitemShut {NoStop}%
\bibitem [{\citenamefont {Magno}, \citenamefont {Grieneisen},\ and\
  \citenamefont {Mar{\'e}e}(2015)}]{magno2015}%
  \BibitemOpen
  \bibfield  {author} {\bibinfo {author} {\bibfnamefont {R.}~\bibnamefont
  {Magno}}, \bibinfo {author} {\bibfnamefont {V.~A.}\ \bibnamefont
  {Grieneisen}}, \ and\ \bibinfo {author} {\bibfnamefont {A.~F.}\ \bibnamefont
  {Mar{\'e}e}},\ }\bibfield  {title} {\enquote {\bibinfo {title} {The
  biophysical nature of cells: potential cell behaviours revealed by analytical
  and computational studies of cell surface mechanics},}\ }\href@noop {}
  {\bibfield  {journal} {\bibinfo  {journal} {BMC biophysics}\ }\textbf
  {\bibinfo {volume} {8}},\ \bibinfo {pages} {8} (\bibinfo {year}
  {2015})}\BibitemShut {NoStop}%
\bibitem [{\citenamefont {Gu}\ \emph {et~al.}(2014)\citenamefont {Gu},
  \citenamefont {Xiao}, \citenamefont {Chen}, \citenamefont {Liu},
  \citenamefont {Huang}, \citenamefont {Li}, \citenamefont {Zhang},\ and\
  \citenamefont {Chen}}]{gu2014robust}%
  \BibitemOpen
  \bibfield  {author} {\bibinfo {author} {\bibfnamefont {J.}~\bibnamefont
  {Gu}}, \bibinfo {author} {\bibfnamefont {P.}~\bibnamefont {Xiao}}, \bibinfo
  {author} {\bibfnamefont {J.}~\bibnamefont {Chen}}, \bibinfo {author}
  {\bibfnamefont {F.}~\bibnamefont {Liu}}, \bibinfo {author} {\bibfnamefont
  {Y.}~\bibnamefont {Huang}}, \bibinfo {author} {\bibfnamefont
  {G.}~\bibnamefont {Li}}, \bibinfo {author} {\bibfnamefont {J.}~\bibnamefont
  {Zhang}}, \ and\ \bibinfo {author} {\bibfnamefont {T.}~\bibnamefont {Chen}},\
  }\bibfield  {title} {\enquote {\bibinfo {title} {Robust preparation of
  superhydrophobic polymer/carbon nanotube hybrid membranes for highly
  effective removal of oils and separation of water-in-oil emulsions},}\
  }\href@noop {} {\bibfield  {journal} {\bibinfo  {journal} {Journal of
  Materials Chemistry A}\ }\textbf {\bibinfo {volume} {2}},\ \bibinfo {pages}
  {15268--15272} (\bibinfo {year} {2014})}\BibitemShut {NoStop}%
\bibitem [{\citenamefont {Singh}\ and\ \citenamefont
  {Singh}(2016)}]{singh2016fabrication}%
  \BibitemOpen
  \bibfield  {author} {\bibinfo {author} {\bibfnamefont {A.}~\bibnamefont
  {Singh}}\ and\ \bibinfo {author} {\bibfnamefont {J.}~\bibnamefont {Singh}},\
  }\bibfield  {title} {\enquote {\bibinfo {title} {Fabrication of zirconia
  based durable superhydrophobic--superoleophilic fabrics using non fluorinated
  materials for oil--water separation and water purification},}\ }\href@noop {}
  {\bibfield  {journal} {\bibinfo  {journal} {RSC advances}\ }\textbf {\bibinfo
  {volume} {6}},\ \bibinfo {pages} {103632--103640} (\bibinfo {year}
  {2016})}\BibitemShut {NoStop}%
\bibitem [{\citenamefont {Wang}\ \emph {et~al.}(2017)\citenamefont {Wang},
  \citenamefont {Xiao}, \citenamefont {Wu}, \citenamefont {Wang}, \citenamefont
  {Du}, \citenamefont {Kong}, \citenamefont {Pan}, \citenamefont {Guan},\ and\
  \citenamefont {Hao}}]{wang2017novel}%
  \BibitemOpen
  \bibfield  {author} {\bibinfo {author} {\bibfnamefont {Z.}~\bibnamefont
  {Wang}}, \bibinfo {author} {\bibfnamefont {C.}~\bibnamefont {Xiao}}, \bibinfo
  {author} {\bibfnamefont {Z.}~\bibnamefont {Wu}}, \bibinfo {author}
  {\bibfnamefont {Y.}~\bibnamefont {Wang}}, \bibinfo {author} {\bibfnamefont
  {X.}~\bibnamefont {Du}}, \bibinfo {author} {\bibfnamefont {W.}~\bibnamefont
  {Kong}}, \bibinfo {author} {\bibfnamefont {D.}~\bibnamefont {Pan}}, \bibinfo
  {author} {\bibfnamefont {G.}~\bibnamefont {Guan}}, \ and\ \bibinfo {author}
  {\bibfnamefont {X.}~\bibnamefont {Hao}},\ }\bibfield  {title} {\enquote
  {\bibinfo {title} {A novel 3d porous modified material with cage-like
  structure: fabrication and its demulsification effect for efficient oil/water
  separation},}\ }\href@noop {} {\bibfield  {journal} {\bibinfo  {journal}
  {Journal of Materials Chemistry A}\ }\textbf {\bibinfo {volume} {5}},\
  \bibinfo {pages} {5895--5904} (\bibinfo {year} {2017})}\BibitemShut {NoStop}%
\bibitem [{\citenamefont {Su}\ \emph {et~al.}(2019)\citenamefont {Su},
  \citenamefont {Liu}, \citenamefont {Li}, \citenamefont {Fang}, \citenamefont
  {He}, \citenamefont {Zhang}, \citenamefont {Li},\ and\ \citenamefont
  {He}}]{su2019rubber}%
  \BibitemOpen
  \bibfield  {author} {\bibinfo {author} {\bibfnamefont {M.}~\bibnamefont
  {Su}}, \bibinfo {author} {\bibfnamefont {Y.}~\bibnamefont {Liu}}, \bibinfo
  {author} {\bibfnamefont {S.}~\bibnamefont {Li}}, \bibinfo {author}
  {\bibfnamefont {Z.}~\bibnamefont {Fang}}, \bibinfo {author} {\bibfnamefont
  {B.}~\bibnamefont {He}}, \bibinfo {author} {\bibfnamefont {Y.}~\bibnamefont
  {Zhang}}, \bibinfo {author} {\bibfnamefont {Y.}~\bibnamefont {Li}}, \ and\
  \bibinfo {author} {\bibfnamefont {P.}~\bibnamefont {He}},\ }\bibfield
  {title} {\enquote {\bibinfo {title} {A rubber-like, underwater
  superoleophobic hydrogel for efficient oil/water separation},}\ }\href@noop
  {} {\bibfield  {journal} {\bibinfo  {journal} {Chemical Engineering Journal}\
  }\textbf {\bibinfo {volume} {361}},\ \bibinfo {pages} {364--372} (\bibinfo
  {year} {2019})}\BibitemShut {NoStop}%
\bibitem [{\citenamefont {Gondal}\ \emph {et~al.}(2014)\citenamefont {Gondal},
  \citenamefont {Sadullah}, \citenamefont {Dastageer}, \citenamefont
  {McKinley}, \citenamefont {Panchanathan},\ and\ \citenamefont
  {Varanasi}}]{gondal2014study}%
  \BibitemOpen
  \bibfield  {author} {\bibinfo {author} {\bibfnamefont {M.~A.}\ \bibnamefont
  {Gondal}}, \bibinfo {author} {\bibfnamefont {M.~S.}\ \bibnamefont
  {Sadullah}}, \bibinfo {author} {\bibfnamefont {M.~A.}\ \bibnamefont
  {Dastageer}}, \bibinfo {author} {\bibfnamefont {G.}~\bibnamefont {McKinley}},
  \bibinfo {author} {\bibfnamefont {D.}~\bibnamefont {Panchanathan}}, \ and\
  \bibinfo {author} {\bibfnamefont {K.~K.}\ \bibnamefont {Varanasi}},\
  }\bibfield  {title} {\enquote {\bibinfo {title} {Study of factors governing
  oil--water separation process using tio2 films prepared by spray deposition
  of nanoparticle dispersions},}\ }\href@noop {} {\bibfield  {journal}
  {\bibinfo  {journal} {ACS applied materials \& interfaces}\ }\textbf
  {\bibinfo {volume} {6}},\ \bibinfo {pages} {13422--13429} (\bibinfo {year}
  {2014})}\BibitemShut {NoStop}%
\bibitem [{\citenamefont {Liu}\ \emph {et~al.}(2016)\citenamefont {Liu},
  \citenamefont {Zhang}, \citenamefont {Yao}, \citenamefont {Zhang},
  \citenamefont {Han},\ and\ \citenamefont {Ren}}]{liu2016facile}%
  \BibitemOpen
  \bibfield  {author} {\bibinfo {author} {\bibfnamefont {Y.}~\bibnamefont
  {Liu}}, \bibinfo {author} {\bibfnamefont {K.}~\bibnamefont {Zhang}}, \bibinfo
  {author} {\bibfnamefont {W.}~\bibnamefont {Yao}}, \bibinfo {author}
  {\bibfnamefont {C.}~\bibnamefont {Zhang}}, \bibinfo {author} {\bibfnamefont
  {Z.}~\bibnamefont {Han}}, \ and\ \bibinfo {author} {\bibfnamefont
  {L.}~\bibnamefont {Ren}},\ }\bibfield  {title} {\enquote {\bibinfo {title} {A
  facile electrodeposition process for the fabrication of superhydrophobic and
  superoleophilic copper mesh for efficient oil--water separation},}\
  }\href@noop {} {\bibfield  {journal} {\bibinfo  {journal} {Industrial \&
  Engineering Chemistry Research}\ }\textbf {\bibinfo {volume} {55}},\ \bibinfo
  {pages} {2704--2712} (\bibinfo {year} {2016})}\BibitemShut {NoStop}%
\bibitem [{\citenamefont {Onda}\ \emph {et~al.}(1996)\citenamefont {Onda},
  \citenamefont {Shibuichi}, \citenamefont {Satoh},\ and\ \citenamefont
  {Tsujii}}]{Kao1995}%
  \BibitemOpen
  \bibfield  {author} {\bibinfo {author} {\bibfnamefont {T.}~\bibnamefont
  {Onda}}, \bibinfo {author} {\bibfnamefont {S.}~\bibnamefont {Shibuichi}},
  \bibinfo {author} {\bibfnamefont {N.}~\bibnamefont {Satoh}}, \ and\ \bibinfo
  {author} {\bibfnamefont {K.}~\bibnamefont {Tsujii}},\ }\bibfield  {title}
  {\enquote {\bibinfo {title} {Super-water-repellent fractal surfaces},}\
  }\href@noop {} {\bibfield  {journal} {\bibinfo  {journal} {Langmuir}\
  }\textbf {\bibinfo {volume} {12}},\ \bibinfo {pages} {2125--2127} (\bibinfo
  {year} {1996})}\BibitemShut {NoStop}%
\bibitem [{\citenamefont {Barab{\'a}si}\ and\ \citenamefont
  {Stanley}(1995)}]{fractalgrowthlivro1995}%
  \BibitemOpen
  \bibfield  {author} {\bibinfo {author} {\bibfnamefont {A.-L.}\ \bibnamefont
  {Barab{\'a}si}}\ and\ \bibinfo {author} {\bibfnamefont {H.~E.}\ \bibnamefont
  {Stanley}},\ }\href@noop {} {\emph {\bibinfo {title} {Fractal concepts in
  surface growth}}}\ (\bibinfo  {publisher} {Cambridge university press},\
  \bibinfo {year} {1995})\BibitemShut {NoStop}%
\end{thebibliography}%

\end{document}